%% file: SG.tex
\documentclass[aps,floats,superscriptaddress]{revtex4}
\usepackage{amsmath}
\usepackage{graphicx}

\def\GroupeEquations#1{\begin{subequations}  #1  \end{subequations}}
	\def\moy#1{\left\langle #1 \right\rangle}

\def\Bigmoy#1{\Bigl\langle #1 \Bigr\rangle}
\def\Im{\hbox{Im}}

\def\enpenche#1{{\sl #1}}

\def\figx#1#2{\includegraphics[width=#1]{#2}}

\def\ie{{\sl i.e. }}
\def\heis{{\text{Heis}}}
\def\bath{{\text{Bath}}}
\begin{document}

\title{Out of equilibrium dynamics of a Quantum Heisenberg Spin Glass}

\author{Giulio Biroli}
\affiliation{Center for Material Theory,
Department of Physics and Astronomy, Rutgers University, Piscataway, NJ
08854-8019 USA}
\author{Olivier Parcollet}
\affiliation{Center for Material Theory,
Department of Physics and Astronomy, Rutgers University, Piscataway, NJ
08854-8019 USA}

\date{\today}
\begin{abstract}
We study the out of equilibrium dynamics  of the infinite range
quantum Heisenberg spin glass model coupled to a thermal relaxation
bath. The $SU (2)$ spin
algebra is generalized to $SU (N)$ and we analyze the large-$N$ limit. 
The model displays a dynamical phase transition between a paramagnetic
and a glassy phase. In the latter,  the system remains out of
equilibrium and displays an aging phenomenon, which we characterize
using both analytical and numerical methods. In the aging regime, 
the quantum fluctuation-dissipation relation is violated and replaced
at very long time by its classical generalization, as in models
involving simple spin algebras studied previously. We also discuss
the effect of a finite coupling to the relaxation baths and their
possible forms. This work completes and justifies previous studies on
this model using a static approach.
\end{abstract}
\maketitle

The study of the non-equilibrium dynamics of classical
glassy systems has been the subject of an intense research in the last decade. A lot of
progress has been made \cite{reviewDYN} using scaling arguments, 
phenomenological approaches and mean field theory. 
One of the major achievement is the theoretical explanation
of the aging phenomena, which is one of the most striking feature
of glassy systems.
The analysis of the out of equilibrium of (classical) mean field
spin glasses has played a major role for several reasons.
It has furnished a framework to understand, interpret and analyze the
experimental results and it has given important predictions on 
the violation and the generalization of the fluctuation dissipation
relation out of equilibrium \cite{CugliandoloKurchanPRL} which has been 
experimentally tested recently \cite{FDTexp}.

Usually, many glassy systems can be analyzed within a classical approach
since they are characterized by transition temperatures at which quantum
mechanical effects are not relevant. Nevertheless, there are also 
interesting cases in which the critical temperature can be lowered
to zero tuning a parameter which controls 
the strength of quantum fluctuations.
This give rise to a quantum critical point at zero temperature \cite{SachdevBook}.
Close to this point, the quantum fluctuations are very important 
and cannot be neglected.
One example which has received much attention recently is the
 insulating magnetic compound ${\mbox{LiHo}_{x}\mbox{Y}_{1-x}
\mbox{F}_{4}}$ which is an experimental realization of an Ising spin glass
in a transverse field \cite{WuBitkoRosenbaumAeppli}.
 Other systems where glassy 
properties in the presence of quantum fluctuations have been observed 
are mixed hydrogen bonded ferro-antiferro electric crystals
\cite{FerroExp},
interacting electron systems \cite{Zvi}, cuprates
like La$_{2-x}$Sr$_x$CuO$_4$ \cite{ChouSG}, 
amorphous insulators \cite{Osheroff1+Osheroff2+Osheroff3}.

The theoretical study of quantum glassy systems has been performed
following two different and complementary routes.  
One dimensional models (like the Random Transverse Ising spin chain)
has been extensively studied and it has been shown that the Griffiths-McCoy 
singularities are very important close to the quantum critical 
point \cite{PapierTheoGriff}.
On the other hand, after the work 
of Bray and Moore \cite{BrayMoore} much attention has been 
focused on infinite dimensional (mean field) models 
\cite{ShuklaSingh,YamamotoIshii,DobroThiru,MillerHuse,kopec,GiamarchiLedoussal,GrempelRozenberg,GrempelRozenberg2,SGletter,SGlong,Isya,rsy,Niri,CugliandoloLozano,CugliandoloPspinQLettre}.
In particular, recently, it has been shown \cite{CugliandoloGrempelDaSilva}
that for the quantum spherical p-spin glass model the quantum fluctuations 
drive the transition toward a {\it first order} quantum phase transition
at low temperature. The same phenomenon has been observed experimentally for 
the  insulating magnetic compound ${\mbox{LiHo}_{x}\mbox{Y}_{1-x}
\mbox{F}_{4}}$ \cite{WuBitkoRosenbaumAeppli}. In \cite{QuantumTAP}
 it has been argued that this phenomenon
is to be expected in a large class of systems.
 
In contrast, the study of real time out-of-equilibrium dynamics of quantum glassy
system is a  recent subject and only very few results are
available at the time of this writing. 
In a first pioneering paper, Cugliandolo and Lozano \cite{CugliandoloLozano}
presented a detailed solution of a quantum version of
the $p-$spin model. They showed how the out of equilibrium behavior 
of classical glassy systems is affected by quantum fluctuations.  
In particular they found that the low temperature glassy phase is 
characterized by the aging phenomenon. In this regime, the fluctuation
dissipation relation is violated and it is generalized to
a form that coincides with the (generalized) classical one. 
This could seem natural since at low frequency 
the quantum fluctuation relation coincides with its $\hbar \rightarrow 0$
limit (for a bosonic system). Indeed it has been shown in 
\cite{ChamonRotorsLettre,ChamonRotors} that for models with simple
commutation relations (particles and rotors)
 the classical nature of the generalized
fluctuation dissipation relation is due to the fact 
the dynamical equations are fixed point of the re-parameterization group
of time transformations and the {\it renormalized} aging dynamics becomes 
classical at the fixed point. The quantum mechanics enters
only as a renormalization of the coefficients of the dynamical equations.

However what happens for models with a non trivial spin algebras, as 
the $SU(N)$ model studied in \cite{SGlong,SGletter}  remained an open
question. 
The study of the out of equilibrium dynamics of this type of quantum glassy systems
is the main aim of this paper.
We will focus on the quantum Heisenberg Spin Glass where the $SU(2)$ spin 
symmetry group is replaced by $SU(N)$ and take the large $N$-limit.
In this model, the spin are true quantum spins, {\sl i.e.} with non
trivial commutation relations, and this
introduces in the problem Berry phases which play an important role
\cite{SachdevBook}.
Recently  a detailed mean field solution using an equilibrium approach
 has been presented  in \cite{SGlong,SGletter}.
The model displays a second order phase transition
at a temperature $T_{eq}$ between a paramagnetic phase and a spin glass
phase, and it is solved by a one-step replica symmetry breaking scheme.
Moreover, using a procedure called ``the marginality condition'',
the existence of a dynamical transition has been predicted at a
temperature $T_{d}>T_{eq}$. 
First introduced and discussed in the quantum case in \cite{GiamarchiLedoussal}, 
this prescription was used in \cite{SGlong,SGletter} 
since it lead to the most acceptable solutions.
Recently, the TAP approach has been fully generalized to quantum systems 
\cite{QuantumTAP}. The relationship between TAP and replica approaches
gives a further hint on why one has to choose the marginal solution 
in the replica method. In fact this solution is related to the marginally 
stable TAP states which have some flat directions around them in the (quantum) 
free energy landscape, on the contrary of all the others which are completely 
stable. Assuming that the quantum out of equilibrium dynamics is dominated 
by the presence of flat directions around the marginally stable TAP states,
 as it happens in the classical case, one finds a more natural justification 
of ``the marginality condition''. However, only a complete dynamical analysis
can fully justify this procedure. The analysis performed in this paper 
of the real time out of equilibrium dynamics, using the
Schwinger-Keldysh close time formalism \cite{Keldysh,RammerSmith}, shows indeed
its correctness. 

This paper is organized as follows : in section \ref{model}, we
present the model and the relaxation bath coupled to it. In section
\ref{EqDynamique}, we present the dynamical large-$N$ equations for
the retarded and Keldysh correlation functions and we explain their
derivation and how to deduce them from the simpler imaginary time
equations.
In section \ref{SectionAging}, we present
both an analytical and a numerical analysis of the dynamical equations.
Numerical evidence for aging and for a generalized
fluctuation-dissipation theorem in the aging regime are presented.
Moreover the analysis of the aging regime justify the ``marginality
condition'' used in previous works \cite{SGletter,SGlong}. Finally, in section
\ref{RoleOfTheBath}, we briefly discuss the effect of a finite coupling to the relaxation bath.

\section{The Heisenberg spin glass and the relaxation bath}\label{model}
The  model considered in this paper is a quantum Heisenberg spin-glass
on a  completely connected lattice of ${\cal  N}$ sites with
quenched disordered couplings $J_{ij}/\sqrt{{\cal N}N}$ which are independent random Gaussian
 variable of zero mean and a variance $\overline{J_{ij}^{2}}/({\cal N}N)=J^{2}_{H}/({\cal N}N)$.
 Each spin is linearly coupled to a thermal
bath. Moreover we generalize the $SU(2)$ 
spin symmetry group to $SU(N)$ and we take the large $N$-limit.
This generalization allows us to obtain a tractable model 
which has still highly non trivial quantum effects and reproduces
qualitatively well the known results for the $SU(2)$ model, 
as far as it has been possible to compare the $N=2$ and the $N=\infty$
cases\cite{SGletter,SGlong,GrempelRozenberg}.
The Hamiltonian reads:
\begin{equation}\label{Hamiltonian}
H = 
\frac{1}{\sqrt{{\cal N}N}} \sum_{i<j} J_{ij} \vec{S}_{i}\cdot
\vec{S}_{j} 
+ \frac{J_{B}}{N\sqrt{\gamma}} \sum_{i\alpha
} \vec{S}_{i}\cdot \vec{s}_{i\alpha  
} + H_{\bath} (\vec{s})
\end{equation}
where the scaling of the spin-spin couplings
and the antiferromagnetic spin-bath coupling has been chosen in such a way to 
obtain a sensible large ${\cal N}$, $N$ limit, i.e. $J^{2}_{H},J_{B} \propto O(1)$.
The first term is the Quantum Heisenberg spin glass Hamiltonian,
the third  and the second  terms represent respectively the thermal
bath of spins $\vec{s}_{i}$ and its coupling to the spins $\vec{S}_{i}$ via the
coupling constant $J_{B}$. Let us now discuss them separately. 

Among the possible representation of the $SU (N)$ spin,  two versions
have been studied \cite{SGletter,SGlong}: the {\sl bosonic model}, in which 
the spin operator $S$ is represented using constrained Schwinger bosons $b$ by 
\GroupeEquations{
\begin{align}\label{DefSchwingerBosons}
S_{\alpha \beta} &=
b^{\dagger}_{\alpha }b_{\beta } - S\delta_{\alpha \beta } \\
\sum_{\alpha =1}^{N}b^{\dagger}_{\alpha }b_{\alpha }&= SN\label{ContrainteDef}
\end{align}
}
and the {\sl fermionic model}, in which the spin operator $S$ is
represented similarly using Abrikosov fermions $f$ by $S_{\alpha \beta}=
f^{\dagger}_{\alpha }f_{\beta } - S \delta_{\alpha \beta }$,
 with the constraint $\sum_{\alpha}
f^{\dagger}_{\alpha }f_{\alpha }= SN$  ($0\leq S\leq 1$).
See \cite{AuerbachBook} for an introduction to this two representations.
The two models are technically very similar but there is an important physical
difference between them : in the fermionic model,  quantum
fluctuations are so strong in the large-$N$ limit 
that the spin glass ordering is destroyed \cite{SachdevYe} (the
critical temperature vanishes when $N$ diverges \cite{SGlong}), whereas in 
the bosonic model, a spin glass phase exists at low temperature
\cite{SGletter,SGlong}. In the following we will focus mainly on the latter 
one and we briefly discuss some results for the former one at the end
of Section \ref{SectionAging}.
In the model we study, the size $S$ of the spin is a fixed, tunable  parameter
 which controls the strength
of quantum fluctuations (since
 $N=\infty$ it is moreover continuous).
Let us emphasize that the $N\rightarrow \infty $ limit is not the
classical limit : as shown in \cite{SGletter,SGlong}, the model is
classical for large $S$ (but not a very low temperature), while it is
``more quantum'' (\ie the quantum fluctuations are more important) at
low $S$ and displays a quantum critical point at $S=0$ where the spin
glass temperature vanishes.

Moreover, it is important to remark that
the real physical object is the spin $S$, not
to  the boson $b$ or the fermion $f$ which should be considered here
more as mathematical tools.
Technically, this leads to a simple $U (1)$ gauge invariance of the 
bosonic or fermionic theory (we can
always multiply the $b$ or the $f$ by a phase) whose consequences will 
be explained later.

Let us now discuss the role of the relaxation bath terms in 
(\ref{Hamiltonian}). Its presence is necessary to allow the energy dissipation. 
It guarantees the relaxation toward equilibrium above the
dynamical transition $T_{d}$ and it is required to obtain an aging
regime below $T_{d}$. In this paper, we are mostly interested in the
$J_{B}\rightarrow 0$ limit, which must always be taken {\bf after} the long
time limit : for example, the dependance of the equilibrium state in
$J_{B}$ is expected to be smooth in this limit, 
although the transient time towards the equilibrium diverges. 
%
%
The bath we have considered in (\ref{Hamiltonian})
is supposed, as usual, to be 
very big and always in equilibrium at a finite temperature
$T=1/\beta$. For further simplification
we take  independent baths   from site to site (labeled by $i$), and the spins $\vec{s}_{i}$
carry an additional degree of freedom $\alpha$, with $1\leq \alpha
\leq N\gamma$ (where $\gamma$ is a constant), which ensures that the bath
is much bigger than the spins it is coupled to.
Moreover we will make the assumption 
of factorized initial condition \cite{QuantumBrownian}. By this we
mean that the initial density matrix is a product of an equilibrium 
density matrix for the bath and an initial density matrix for the system.
Since $b$ (or $f$) is not the physical object, the bath should respect
the $U (1)$ invariance, {\sl i.e.} it must couple to two $b$'s and we
do not consider baths coupling linearly to $b$ in this paper.
Of course many different choices are possible. We first 
consider a ``generic'' bath $H_{bath} (\vec{s})$ of
interacting spins $\vec{s}$ and expand the Keldysh effective action
for $S$ at second order in the coupling constant
$J_{B}$ (the first order vanishes).
In the dynamics, the bath will then appear only
through its susceptibility $\chi_{0}$ (see section \ref{EqDynamique}). This approach
is appropriate when we only consider the bath as a device 
to provide thermalization. One could wonder how correct
is to study a low temperature glassy phase using a perturbative treatment 
of the coupling to the environmental heat bath. In Section
\ref{RoleOfTheBath}
we will show that this is not a limitation and we will briefly discuss
the simplest type of spin bath which will turns out to be a 
Kondo bath. 

\section{The dynamical equations}\label{EqDynamique}
In order to study the real time dynamics, we use the 
Keldysh method \cite{Keldysh} : the time
evolution operator is written as a path integral over 2 times $t_{+}$
and $t_{-}$, running from $0$ to infinity forward and backward
respectively.
 In the classical limit, this method reduces to the
Martin-Siggia-Rose-DeDominicis-Janssen formalism (See Appendix C of
\cite{CugliandoloLozano}). 
We take an infinite temperature initial condition at $t=0$ and do a
instantaneous quench to temperature $T$. This means that the initial
density matrix for the system is simply the identity operator.
As a consequence the initial density matrix, which is a product of 
the equilibrium density matrix of the bath and the infinite
temperature density matrix of the system, does not depends on the
disordered couplings. Therefore, as in the classical case,
there is no need to introduce replicas to compute disorder-averaged quantities.

The quantities that we want to compute are the averaged response and correlation
of the spin, which are defined as (the spins are on the same site) :  
\GroupeEquations{
\begin{align}\label{DefinitionRCSpin}
R_{S} (t,t') &\equiv  \frac{i}{N^{2}}  \theta (t-t')
\overline{
 \Bigmoy{
\vec{S} (t)\cdot\vec{S} (t')-\vec{S} (t')\cdot\vec{S} (t)
} 
}= \frac{i}{N^{2}} \sum_{\alpha \beta } \theta (t-t')
\overline{
 \Bigmoy{\left[
S_{\alpha \beta } (t),S_{\beta \alpha } (t')
 \right]}} 
\nonumber \\
C_{S}(t,t') &\equiv  
\frac{1}{2N^{2}}
\overline{
\Bigmoy{\vec{S} (t) \cdot \vec{S} (t') + \vec{S} (t') \cdot 
\vec{S} (t)}
}=\frac{1}{2N^{2}} \sum_{\alpha \beta }
\overline{
\Bigmoy{S_{\alpha \beta } (t) S_{\beta \alpha } (t') + S_{\alpha \beta
} (t') S_{\beta \alpha } (t)}
}
\end{align}
}
where the bar denotes the average over disorder and the brackets denote
the ``Keldysh average'' i.e. the Hamiltonian evolution of the quantity
starting from the initial condition at $t=0$. Moreover, in this paper,
we take $\hbar =1$.
 
In the following, we first derive the Keldysh action,
we average over disorder and we take the large-$N$ limit; then we recall 
the so-called ``Larkin-Ovchinnikov representation'' and we express the
dynamical large-$N$ equations in their final form using the retarded
and Keldysh functions of the bosons $R$ and $K$ from which one can obtain
 $R_{S}$ and $C_{S}$. It is not possible to
obtain tractable equations for the physical quantities $R_{S}$ and
$C_{S}$ directly, contrary to rotors \cite{ChamonRotors} or $p-$spins
\cite{CugliandoloLozano} models, and that makes the problem more complicated
\cite{SGletter,SGlong}.

We start using a Keldysh action defined on the double contour :
\begin{equation}\label{DefKeldyshAction}
{\cal S} =  S_{B}^{+} - S_{B}^{-} - i 
\sum_{a=+,-} {a} \int_{0}^{\infty} dt 
\left[
\sum_{\moy{i,j}} \frac{J_{ij}}{\sqrt{{\cal N}N}} \vec{S}^{a}_{i} (t)\cdot\vec{S}^{a}_{j} (t) 
 +  
\frac{J_{B}}{N\sqrt{\gamma}} \sum_{i\alpha } \vec{S}^{a}_{i} (t)\cdot \vec{s}^{a}_{i\alpha
} (t) + H_{\text{bath}} (\vec{s}^{a}_{i} (t)) 
 \right]
\end{equation}
In this expression, $a=\pm$ denotes the upper/lower contour,
$S_{B}^{\pm}$ are the Berry phase on the upper and lower contours
\cite{Note_Keldysh}.
After the average over the disorder we take a saddle point over
the number of sites ${\cal N}$. Hence, we get a self-consistent
problem (some scalar products have been explicitly written with
$SU (N)$ indices for clarity) : 
\begin{multline}\label{KeldyshAction1Site}
{\cal S_{2}} = S_{B}^{+} - S_{B}^{-} -  
\sum_{a,b=+,-} ab
 \iint_{0}^{\infty}  dt dt' 
 \frac{J^{2}_{H}}{2N} 
\sum_{\alpha \beta \gamma \delta }
{S}^{a}_{\alpha \beta }
{S}^{b}_{\gamma \delta } (t') 
\moy{{S}^{a}_{\beta \alpha } (t){S}^{b}_{\delta \gamma }
(t')}_{\cal S_{2}}
  \\ 
-i \sum_{a=+,-} a \int_{0}^{\infty}  dt 
\frac{J_{B}}{N\sqrt{\gamma}} 
\left(
\sum_{\alpha } \vec{S}^{a} (t)\cdot \vec{s}^{a}_{\alpha
} (t) + H_{\text{bath}} \bigl (\vec{s}^{a} (t) \bigr) 
\right)
\end{multline}
where $\moy{\cdot}_{\cal  S_{2}}$
 means the average over the single site action (\ref{KeldyshAction1Site}).

At this stage, it is useful to define the spin correlation  functions
in the so called  ``$\pm$ representation''  :
\begin{align}\label{defSpinGreenFunctionPM}
\chi^{S}_{ab} (t,t')&\equiv \frac{1}{N^{2}}
\moy{{\cal  T}\vec{S}^{a} (t) \cdot\vec{S}^{b}(t') }_{\cal S}
= \frac{1}{N^{2}}
\sum_{\alpha \beta }
\moy{{\cal  T}S^{a}_{\alpha \beta } (t) S^{b}_{\beta\alpha} }_{\cal S}\\
\chi^{0}_{ab} (t,t')&\equiv \frac{1}{N^{2}}
\moy{{\cal  T}\vec{s}^{a} (t)\cdot  \vec{s}^{b} (t')}_{\text{Bath}}
\end{align}
where $a,b=\pm$ are the contour indices, and ${\cal T}$ is the time ordering on the
double contour,  $\moy{\cdot}_{\text{Bath}}$ denotes the average with respect
to the bath.
Now, using explicitly the $SU (N)$ invariance of the theory,
integrating out the bath, and expanding to second order in $J_{B}$ we get
an action only for the spins (summation  over $\alpha $ and $\beta $
is implicit) : 
\begin{equation}\label{KeldyshAction2}
{\cal S} = S_{B}^{+} - S_{B}^{-} -  
\frac{1}{2N}
\sum_{a,b} ab
 \iint_{0}^{\infty}  dt dt' 
{S}^{a}_{\alpha \beta } (t){S}^{b}_{\beta\alpha } (t')
\Bigl (
 J^{2}_{H} \chi^{S}_{ab} (t,t') 
+
J_{B}^{2}\chi^{0}_{ab} (t,t')
\Bigr )
\end{equation}

\begin{figure}[ht]
\[
\figx{14cm}{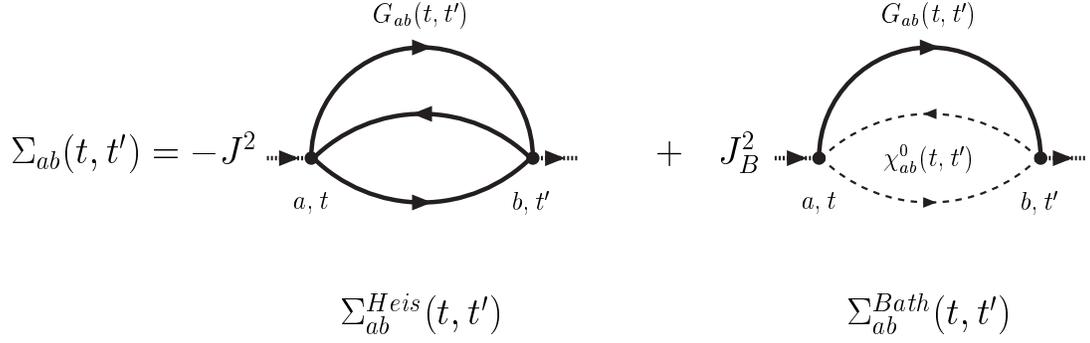}
\]
\caption{\label{Diagrams}
\sl Feynman diagrams for the self energy : the solid line is the
bosonic Green function, the dashed lines represents the $\chi^{0}$ of the bath (we
take a product of two fermionic functions, see text).}
\end{figure}

Until now, the derivation is correct for any value of $N$ and in particular
for $N=2$. The great technical advantage of the large-$N$ limit becomes
manifest if one considers in detail the self-consistent single site
problem. In fact, because of the presence of the Berry phase, 
the single site measure defined by the action (\ref{KeldyshAction2})
is far from being simple. Indeed the single-site functional integral 
cannot be performed and as a consequence it is not possible to obtain a closed 
equation for the spin-spin correlation function.
The large-$N$ limit simplifies
the single-site measure and gives  a set of 
closed equation on the two-point functions. Using the Schwinger
bosons,  the Berry phase contribution to (\ref{KeldyshAction2}) reads:
\begin{equation}\label{Berrybos}
{\cal  S}_{B}=-\sum_{a, \alpha }{a}\int_{0}^{\infty }dt
b^{a\dagger}_{\alpha }\partial_{t}b^{a }_{\alpha }
\end{equation}
while the other part of the action can be obtained simply replacing
${S}^{a}_{\alpha \beta }$ with its expression in terms of bosons. For a finite $N$ the problems
remains still very complicated since one has to integrate only on 
bosonic fields respecting the constraint (\ref{ContrainteDef}). 
Whereas in the large-$N$ limit, which we shall study in the following,
 the sum $\sum_{\alpha }b_{\alpha }^{\dagger
}b_{\alpha }/N$ does not fluctuate and this greatly simplifies the analysis.
In particular
we can 
obtain the saddle point equations on the bosonic Green functions, which in
the $\pm$  representation are defined as : 
\begin{equation}\label{defGreenFunctionPM}
G_{ab} (t,t') \equiv  - i \moy{ {\cal T} b_{a} (t) b^{\dagger}_{b} (t')}_{\cal S}
\end{equation}
The computation can be done explicitly, using the same decoupling as
in the imaginary time equilibrium computation, as explained in
\cite{SGlong}. Here we present a faster derivation, noting
that the same diagrams for the self
energy derived in the imaginary time computation appears
in the Keldysh formalism :
a first diagram, corresponding to the spin glass interaction itself, 
and a second one, corresponding to the coupling to the relaxation bath 
(See Figure \ref{Diagrams}).
Thus using Feynman rules in the $\pm$ representation, we find
immediately (the factors can be checked against Matsubara computations
: see appendix A) : 
\begin{equation}\label{EqsGeneralPM}
\Sigma_{ab} (t,t') = - \underbrace{J^{2} ab G_{ab}^{2}(t,t')
G_{ba} (t',t)}_{\displaystyle \Sigma^{\heis}_{ab} (t,t')}
+
\underbrace{\mathstrut J^{2}_{B} ab \chi_{ab}^{0} (t,t') G_{ab} (t,t') 
}_{\displaystyle \Sigma^{\bath}_{ab} (t,t')}
\end{equation}

We see that the bath only enters though its susceptibility
$\chi^{0}$. For simplification, in the following, we take a specific
form for the susceptibility of the bath 
$\chi^{0}_{ab} (t,t') =G_{ab}^{0} (t,t')G_{ba}^{0} (t',t)$
where $G^{0}$ is the Green function of free fermions with a Lorentzian 
density of states at half filling (See section IV for a discussion 
on the relaxation bath and a justification of this formula).

To simplify the analysis of the dynamical equations, it is useful
to write them in a different way, using 
the so-called  ``Larkin-Ovchinnikov
representation'' (LO) of the equations, in which the (matrix) Green
function is given by 
\begin{equation}\label{DefLO}
G = \frac{1}{2} 
\begin{pmatrix}
1 & 1\\
1 & -1
\end{pmatrix}
\begin{pmatrix}
G_{++} & G_{+-}\\
G_{-+} & G_{--}
\end{pmatrix}
\begin{pmatrix}
1 & 1\\
-1 & 1
\end{pmatrix}
=
\begin{pmatrix}
R & K\\
0 & A
\end{pmatrix}
\end{equation}
where the retarded (response), advanced
and Keldysh (correlation) two-times Green functions are defined by : 
\GroupeEquations{
\begin{align}\label{DefGreenFunctionLO}
R (t,t') &\equiv  - i \theta (t-t')\moy{\bigl[ b (t), b^{\dagger}
(t')\bigr ]} = G_{++} (t,t') - G_{+-} (t,t') \\
A (t,t') &\equiv  + i \theta (t'-t)\moy{\bigl[ b (t), b^{\dagger}
(t')\bigr ]}  = G_{++} (t,t') - G_{-+} (t,t')  \\
K (t,t') &\equiv  - i \moy{\bigl\{ b (t), b^{\dagger} (t')\bigr
\}} = G_{++} (t,t') + G_{--} (t,t') 
\end{align}
}
Note that our convention for $R$ differs from the one used for
$R_{S}$. 
Moreover, we will use the relations : 
\begin{align}\label{RelationsPHsym}
A (t,t') &=R (t',t)^{*} \\
K (t,t') &=- K (t',t)^{*}
\end{align}
to eliminate $A$ and restrict ourselves to $t>t'$.
The (LO) representation is simpler because it uses the
relation $G_{++}+G_{--}=G_{-+}+G_{+-}$ to reduce the number of
functions (just $R$ and $K$, after that $A$ is eliminated using
(\ref{RelationsPHsym}) and because it makes the causality
of the equations explicit.
In the (LO) representation the
Keldysh indices structure of the vertices are particularly simple : the 2-leg
vertex is $\delta_{ab}$ \cite{RammerSmith}, which leads to a simple
Dyson equation  $G^{-1} = G^{-1}_{0} - \Sigma$ (where the inverse are
just matricial inverses), and the four-leg vertex used in
(\ref{EqsGeneralPM}) is $(1\otimes \sigma_{x} + \sigma_{x} \otimes 1)/2$,
where $1$ is the $2\times 2$ identity matrix and  $\sigma_{x}$ is the
usual Pauli matrix. Remarkably, this vertex factor is fully symmetric in the
Keldysh space.
Hence, a quick way to switch to the (LO) representation is to recompute the
Feynman diagrams (See Figure (\ref{Diagrams})) with the (LO) Feynman
rules (although a direct computation using just the definitions is
possible but more tedious). 
After these manipulations,  we finally obtain the main  equations of
our paper (for $t>t'$) :  
\GroupeEquations{\label{EqsGenerales}
\begin{itemize}
\item The Dyson equations ((\ref{DysonK2}) is rewritten to involve  only functions for
$t>t'$): 
\begin{align}
 i\partial_{t} R (t,t') &=  \delta (t-t')
+ \int_{t'}^{t} du \,\, \Sigma_{R} (t,u) R (u,t')
\label{DysonR}
\\
\label{DysonK}
i\partial_{t} K (t,t') &= 
\int_{0}^{t} du \,\,  \Sigma_{R} (t,u)K (u,t') 
+
\int_{0}^{t'} du \,\,
\Sigma_{K} (t,u) R^{*}(t',u)
\\
&= 
\int_{t'}^{t} du \,\,  \Sigma_{R} (t,u)K (u,t') 
+
\int_{0}^{t'} du \,\,
\biggl (
\Sigma_{K} (t,u) R^{*}(t',u) - \Sigma_{R}
(t,u) K^{*}(t',u)
\label{DysonK2}
\biggr )
\end{align}
\item The boundary conditions, which derive respectively from (\ref{ContrainteDef}) and the
commutation relations of the boson :
\begin{align}\label{Boundary}
K (t,t) &= -i (2S + 1)
\\
\lim_{t\rightarrow (t')^{ +}}R (t,t') &= -i 
\end{align}
\item The self-energy in {\sl LO} representation (these formulas are local in
times, so the argument $(t,t')$ has been omitted for clarity) :
\begin{equation}
\Sigma \equiv \Sigma^{\heis} +\Sigma^{\bath}
\end{equation}
\def\argu{(t,t')}
\def\argu{}
\begin{align}
\Sigma_{R}^{\heis}\argu &\equiv  
- \frac{J_{H}^{2}}{4}
\biggl(
 \Bigl( \big |R\argu \big |^{2} - \big |K\argu\big |^{2} \Bigr)R\argu
+ 
 \Bigl( R^{*}\argu K\argu - K^{*}\argu R\argu \Bigr)K\argu
\biggr)
\label{SigmaLOR}
\\
\Sigma_{K}^{\heis}\argu &\equiv  
- \frac{J_{H}^{2}}{4}
\biggl(
 \Bigl( \big |R\argu \big |^{2} - \big |K\argu\big |^{2} \Bigr)K\argu
+ 
 \Bigl( R^{*}\argu K\argu - K^{*}\argu R\argu \Bigr)R\argu
\biggr)
\label{SigmaLOK}
\end{align}
\begin{align}
\Sigma_{R}^{\bath} \argu  &\equiv  
 \frac{J_{B}^{2}}{4}
\biggl(
 \Bigl( \big |R_{0}\argu \big |^{2} - \big |K_{0}\argu\big |^{2} \Bigr)R\argu
+ 
 \Bigl( R^{*}_{0}\argu K_{0}\argu - K^{*}_{0}\argu R_{0}\argu \Bigr)K\argu
\biggr)
\label{SigmaLORb}
\\
\Sigma_{K}^{\bath}\argu &\equiv  
 \frac{J_{B}^{2}}{4}
\biggl(
 \Bigl( \big |R_{0}\argu \big |^{2} - \big |K_{0}\argu\big |^{2} \Bigr)K\argu
+ 
 \Bigl( R^{*}_{0}\argu K_{0}\argu - K^{*}_{0}\argu R_{0}\argu \Bigr)R\argu
\biggr)
\label{SigmaLOKb}
\end{align}
\end{itemize}
}
where $R_{0}$ and $K_{0}$ are defined in the same way that $R$ and $K$
 starting from $G^{0}_{a,b}$.
This expression emphasizes the very similar structure of the two terms
in the self-energy :  more generally,  the bath term reads
$\Sigma_{R}^{\bath} \propto  \chi^{0}_{K}  R  + \chi^{0}_{R} K $ and 
$\Sigma_{K}^{\bath} \propto  \chi^{0}_{R}  R  + \chi^{0}_{K} K$, where
$\chi^{0}_{R}$ and $\chi^{0}_{K}$ are the retarded and the Keldysh
part of $\chi^{0}$ respectively.

Finally, we now derive the expression for the response and correlation
functions for
spins defined in (\ref{DefinitionRCSpin}). In the $N \rightarrow \infty $ 
limit, $C_{S}$ and  $R_{S}$  can be easily computed from $R$ and $K$
 using the relations :
\GroupeEquations{
\begin{align}\label{RetardedSpin}
R_{S} (t,t') 
&= \frac{i}{N^{2}}\sum_{\alpha \beta }\nonumber
\Bigmoy{S^{+}_{\alpha \beta } (t)
 \left(S^{+}_{\beta \alpha } (0) - S^{-}_{\beta \alpha } (0) \right)} 
\\
&=\Im
 \left(
   K (t,t') R^{*} (t,t')
 \right)
\\
\nonumber
C_{S}(t,t') &=
\frac{1}{N^{2}}\sum_{\alpha \beta }
 \Bigmoy{S^{+}_{\alpha \beta } (t) S^{+}_{\beta \alpha } (0) -
S^{-}_{\alpha \beta } (t) S^{-}_{\beta \alpha } (0)} 
\\
&= \frac{1}{4}
 \Bigl(
   |K (t,t')|^{2} - |R (t,t')|^{2} - |R (t',t)|^{2}  
 \Bigr)
\end{align}
}
To derive (\ref{RetardedSpin}), we used
Eqs. (\ref{DefSchwingerBosons},\ref{ContrainteDef}), the $SU (N)$
invariance, the
limit $N\rightarrow \infty $ (to drop subdominant terms), and the
relations between $G_{a,b}$ and $R,K,A$ that invert
(\ref{DefGreenFunctionLO}).

The equations (\ref{EqsGenerales}) describe the dynamics of the Quantum Heisenberg spin
glass in the large-$N$ limit, coupled to the bath.
In the following sections, we  present an analysis of these equations
both in the paramagnetic regime and in the aging regime,
together with some results extracted from a numerical solution of this
systems.
Let us first make a few preliminary remarks :
\begin{itemize}
\item [\enpenche{i)}] Contrary to the quantum $p-$spin problem studied
in \cite{CugliandoloLozano}, there is no need here for
a Lagrange parameter associated to the constraint. This is due to the
fact that the constraint (\ref{ContrainteDef}) commutes with the Hamiltonian and hence is
conserved in the time evolution. Indeed one can check that
(\ref{SigmaLOR},\ref{SigmaLOK},\ref{SigmaLORb},\ref{SigmaLOKb}) and (\ref{DysonR},\ref{DysonK2}) imply: 
\begin{equation}\label{CheckBoundaryConserved}
\frac{d K (t,t)}{dt} \propto \Im \frac{\partial}{\partial t} K \left.(t,t') \right|_{t'=t} = 0
\end{equation}
Thus if the boundary condition is verified for $t=t'=0$, it will
propagate at all later times.

Formally, one can introduce such a parameter $\lambda $ in the
equation, by replacing $i\partial_{t}$ by  $i\partial_{t}+ \lambda$,
but it can be removed using a $U(1)$ gauge transformation
$[K(t,t'),R(t,t')]\rightarrow [\tilde{K}(t,t'),\tilde{R}(t,t')]=[e^{-i\lambda (t-t')}K(t,t'),
e^{-i\lambda (t-t')}K(t,t')]$ : if $K,R$ are a solution of the
equations with $\lambda$,$[\tilde{K},\tilde{R}]$ are a solution with $\lambda=0$. 
This symmetry comes from the fact that we represented the spin (the
physical object) with the bosons (a mathematical tool) and that 
the bath couples to the spin and not to the boson, and thus can not
break the symmetry; this has an important consequence (see {\sl ii}).
 \item  [\enpenche{ii)}]
In {\bf equilibrium}, the {\bf spin} response and correlation functions are related by 
the {\sl quantum fluctuation dissipation relation} QFDR:
\begin{equation}\label{QFDR}
R_{S}(\tau )=i\theta (\tau )\int_{-\infty }^{+\infty }\frac{d\omega }{\pi }\exp (-i\omega \tau )
\tanh \left(\frac{\beta \omega }{2} \right)C_{S}(\tau )
\end{equation}
It is important to notice that instead the boson response and
correlation functions $R$ and $K$
are not in principle related by this QFDR. Technically, this is due to the $U (1)$
invariance explained above : if $[K(t,t'),R(t,t')]$ satisfies QFDR
(\ref{QFDR}), $[\tilde{K},\tilde{R}]$ will not in general. Imposing
QFDR for the boson demands that $\lambda $ takes a precise value
$\lambda_{0}$.
 However, this is not a problem since the only physical objects are
 the spin response and correlation functions.
When using the Matsubara imaginary time formalism, one does not face
this difficulty, since 
one automatically requires the QFDR to be satisfied, because of the
$\beta-$periodicity of the imaginary time boson Green function. Thus, when
we will analytically continue our equations in imaginary time to compare to
Matsubara computations (See Appendix \ref{AppendiceMatsubara} ), we will have to reintroduce
$\lambda_{0}$.
\item [\enpenche{iii)}] The presence of the thermal bath is clearly
required : 
for $J_{B}=0$, the solution of the equations has the property 
 $K (t,t') = (2S+1)R (t,t')$ for all $t,t'$ (this can be check order
by order in the coupling constants $J_{H}$ and $J_{B}$, using
(\ref{DysonR},\ref{DysonK2},\ref{SigmaLOR},\ref{SigmaLOK},\ref{SigmaLORb},\ref{SigmaLOKb})) and it is clearly
incorrect (for example, it can not satisfy the high-temperature limit 
of the fluctuation-dissipation relation).
\item [\enpenche{iv)}] We can immediately generalize these equations in the fermionic
case by changing $S\rightarrow -S, J_{H}^{2}\rightarrow
-J_{H}^{2}$ ($S$ is the size of the ``fermionic'' spin, and the sign
change in front of the coupling constant comes from the fact that
there is now a fermion loop in the diagram).
 We will see in Section \ref{SectionAging}  that this simple change
leads to the disappearance of the aging phenomena, as expected \cite{SachdevYe}.
\end{itemize}

\section{Solutions of the dynamical equations}\label{SectionAging}
In this section, we present the solution of the dynamical equations
(\ref{EqsGenerales}) using both numerical and analytical results.
Indeed, these integro-differential equations are causal, so one can
construct the solution step by step in time. This property is very
general (See \cite{CugliandoloLozano} for another example) and it is
the basis for the numerical algorithms, although in this problem
some new technical refinements are needed in order to compute an
accurate solution at a reasonable cost in computational time (see the appendix
\ref{AppendiceNumerique} for a detailed discussion).
The numerical solution shows that the model has a dynamical phase transition
at a temperature $T_{d}(S,J_{B})$ between a paramagnetic phase ($T>T_{d}$) 
and a glassy phase ($T<T_{d}$), as expected on general grounds and predicted
in \cite{SGletter,SGlong}. At high temperature, the system equilibrates 
inside the paramagnetic state: after a transient time all the two-time
quantities become time-translation invariant (TTI) and the quantum
fluctuation dissipation relation QFDR holds.
Instead at low temperature the system never equilibrates on finite
timescales 
\cite{Note_FiniteTimescales}
and one can identifies two different time
sectors on which the two-time functions evolve: when $t'$ is large 
but the difference $t-t'$ is of the order of one and very small
compared to $t'$ the system seems to be equilibrated (the 
QFDR is approximatively verified and all the one time
quantities, as the energy or the Edwards-Anderson parameter,
have almost converged to their asymptotic values). However, on larger
timescales, when $t-t'$ is of the same order of $t'$, an extremely
slow dynamics sets in. In this regime the QFDR is violated and the
aging phenomenon appears \cite{CugliandoloLozano,reviewDYN,CugliandoloKurchanPRL}.
We remark that if one takes the long time limit and then 
sets the coupling to the bath to zero then
the dynamical solution gets back to the equilibrium 
for $T>T_{d}$ only. For $T<T_{d}$ the system never reaches a stationary 
solution. However, the pseudo-equilibrium solution reached in the 
time sector $t-t'\sim O(1)<<t'$ can be also obtained by a pure 
static computation using the marginality prescription \cite{SGlong}.
In the following, we will take $J_{B}>0$ ($J_{B}=1$ for numerical
computation).
 In section IV,  we will discuss what happens changing the value of $J_{B}$.

\subsection{Equilibration into the paramagnetic state}\label{para}
\begin{figure}[ht]
\[
\figx{14cm}{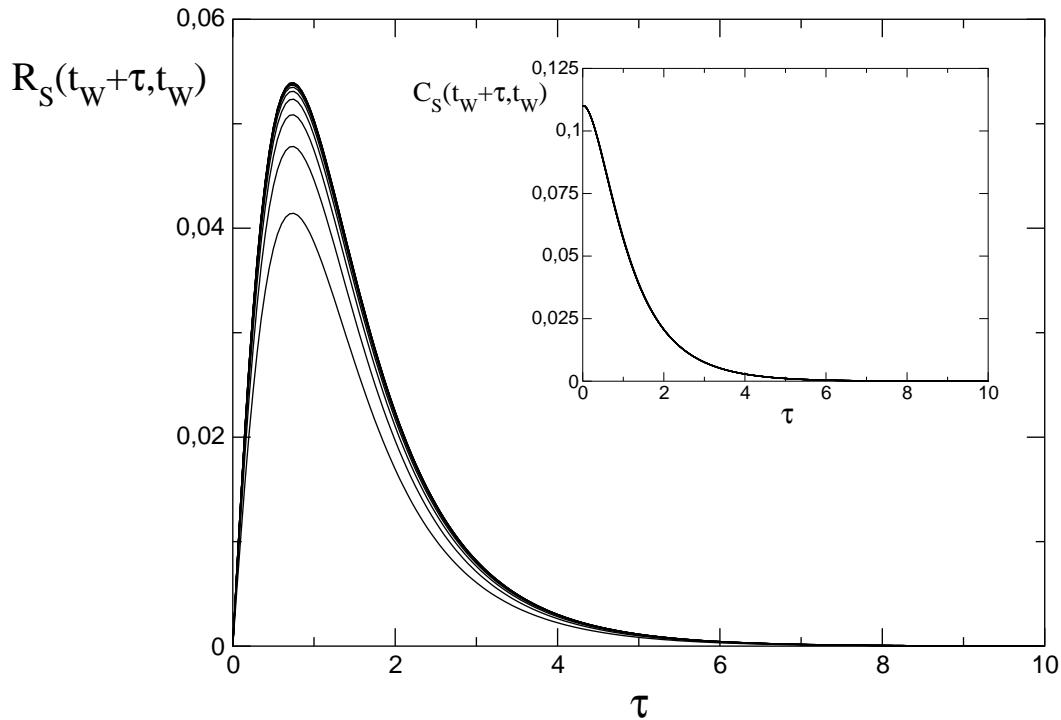}
\]
\caption{\label{para.fig}
\sl Spin response as a function of $\tau $ for
$t_{w}=1.5,2,2.5,3,3.5,4,4.5,5,10,15,20,25$ (from bottom to top
($t_{max}=100$)), $J_{B}=J_{H}=1$, $S=1$ and $T=10$ (inside the
paramagnetic phase). After a transient time the response 
converges to a stationary equilibrium regime. Inset: spin correlation 
as a function of $\tau $ for the same parameters. 
}
\end{figure}

At high temperature, the numerical solution shows,
as expected, that the system equilibrates into the paramagnetic state after 
a transient time $t_{eq}$: for $t,t'>>t_{eq}$ the response and
correlation becomes a function of $\tau =t-t'$ only
and they are related by the QFDR  (see Fig. \ref{para.fig}).
This is indeed what we obtain from
the analysis of Eqs. (\ref{DysonR},\ref{DysonK2}) in the limit 
$t,t'\rightarrow \infty $ with $\tau =t-t'$ fixed. 
In this limit the
equation on $K$ and $R$  can be easily written in Fourier space (we
reintroduce the $\lambda $ term, in agreement with the discussion at
the end of section \ref{EqDynamique}):
\GroupeEquations{
\begin{eqnarray}\label{eqpara1}
R^{-1}(\omega )&=&\omega + \lambda  -\Sigma _{R}(\omega )\\
K(\omega )&=& \Sigma _{K}(\omega ) |R(\omega )|^{2} 
\label{eqpara2}
\end{eqnarray}
}
\begin{figure}[ht]
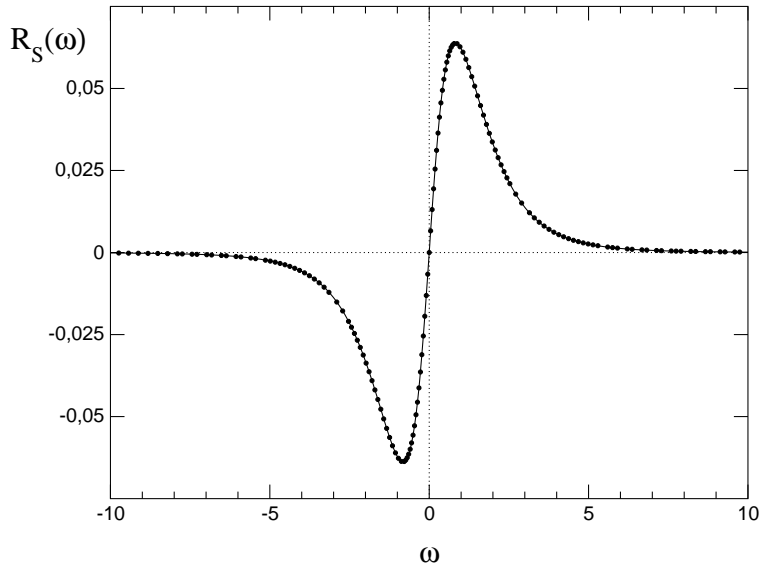

\[
\figx{10cm}{ParaFDT.eps}
\]
\caption{\label{ParaFDT.fig}
\sl Fourier transform of the spin response function (dots) compared to 
its expression (continuous line) computed by the QFDR from the correlation for
 $J_{B}=J_{H}=1$, $S=1$, $t_{W}>10$ and $T=10$ (inside the paramagnetic phase). 
The excellent agreement shows that the system is fully equilibrated.
}
\end{figure}
As in \cite{CugliandoloLozano}, it is possible to show order by order in
perturbation theory that these equations admit 
a solution such that the spin correlation and response functions satisfy QFDR.
In Figure \ref{para.fig},  we plot the spin correlation function 
$C_{S}(t_{w}+\tau ,t_{w})$  and the response function $R_{S}(t_{w}+\tau ,t_{w})$ 
as a function of $\tau $ for different $t_{w}$ for $J_{B}=1$, $J_{H}=1$
and $T=10$. These figures represents the typical behavior of $C_{S}$ and 
$R_{S}$ in the paramagnetic phase: after a short transient time the
functions become TTI (they do not depend on $t_{w}$ anymore)
and they decay quickly as a function of $\tau $.
Moreover, in Figure \ref{ParaFDT.fig}, we plot the retarded function
computed directly from  the numerical solution and from the correlation
function using the QDFR. The excellent agreement shows that QFDR is
satisfied and the system has relaxed to equilibrium. \\
This asymptotic solution represent the equilibrium dynamics inside the
paramagnetic state and it exist only above $T_{d}$. At $T=T_{d}$ 
the equilibration time diverges and remains infinite in all the low
temperature phase ($T<T_{d}$), as clearly indicated by the numerical
solution (see Fig. \ref{CAG.fig}). The study of this regime is the
subject of the next subsections.

In principle,  one would like to take a small $J_{B}$, so that it allows
the system to relax but does not change the value of the paramagnetic
state. However,  the relaxation time diverges when $J_{B}$ goes to 0,
even in the paramagnetic state, and this prevent the numerical program
to converge towards the solution in a reasonable amount of time. 
We found that $J_{B}=1$ is a good compromise. Indeed the imaginary time
computation shows that the results at $J_{B}=0$ are close to $J_{B}=1$.
This relatively big value reflects the fact our bath, coupling to the
spin degrees of freedom, is not very efficient. A more precise
discussion will be given in Section \ref{RoleOfTheBath}.

\subsection{General properties of the glassy dynamics}\label{properties}
The numerical results and the analytical analysis of the dynamical equations indicates
 that the system remains always out of equilibrium at low temperature ($T<T_{d}$). 
In the following we present the Ansatz which gives the asymptotic solution in the 
glassy regime and we compare it to the numerical results obtained integrating the dynamical
equation numerically. This Ansatz is a slight generalization of 
the one introduced by Cugliandolo and Lozano \cite{CugliandoloLozano} for 
quantum glassy systems, which is itself a generalization of the one discovered
by Cugliandolo and Kurchan for classical glassy systems \cite{CugliandoloKurchanPRL}.
\subsubsection{The weak-ergodicity breaking and the weak long-term
memory Ansatz}\label{Ansatz}
In the long time limit ($t,t'>>1$) we make the following Ansatz for the behavior
of the bosonic correlation and response function \cite{CugliandoloLozano}:
\GroupeEquations{
\begin{align}\label{ansatz1}
K(t,t')&\simeq
\left[K_{ST}(t-t')+K_{AG}\left(\frac{h(t')}{h(t)}\right)\right]e^{-i\lambda
(t-t')}\\ 
R(t,t')&\simeq  
\left[R_{ST}(t-t')+\frac{h'(t')}{h(t)}R_{AG}\left(\frac{h(t')}{h(t)}\right)\right]e^{-i\lambda
(t-t')}\label{ansatz2}\\ 
\lim_{t\rightarrow \infty }K_{ST}(t)&=0\\
K_{ST}(0)&=-i(2S+1)+ig\\
K_{AG}(0)&=0\\
K_{AG}(1)&=-ig \label{boundaryansatz}
\end{align}
}
where $h(t)$ is an increasing function of $t$,
 $h'(t)$ is the first derivative of $h(t)$, $g$ is a real number 
\cite{Note_EA} 
and $g^{2}/4$ is equal to the Edwards-Anderson parameter $q_{EA}$. Note
the physics hidden in this Ansatz: 
in the time regime in which $t,t'$ are large but their difference remains finite
 (called TTI-regime in the following) the system seems to have reached a stationary state,
however on a timescale diverging with $t,t'$ ($t,t'$ are large but the ratio
$h(t')/h(t)$ remains finite) there is a secondary evolution called aging
\cite{reviewDYN}. This unveils the interpretation of $h(t)$ as 
a self-generated effective time scale.

Moreover we notice that if this scenario
is realized for the bosonic correlation and response functions then it
will be also realized for the spin correlation and response functions
and for the self-energies. 
The presence of the oscillating exponential is a 
slight generalization with respect to \cite{CugliandoloLozano} and it
could be gauged away, as discussed before. Moreover, when one 
compute the {\sl spin} correlation and response functions these
exponentials cancel.

The second key ingredient that makes the asymptotic problem tractable is 
the assumption of {\it the weak long-term memory property} 
\cite{CugliandoloLozano,CugliandoloKurchanPRL} that allows one to decouple 
the transitory regime from the asymptotic one. In fact the dynamical equations
contains explicitly memory terms which couple all the timescales. So, how
can one analyze the asymptotic regime without solving the complete problem?
Within the weak long-term memory scenario the (linear) response to a finite time perturbation
vanishes  in the long time limit ($t,t'\rightarrow \infty $), whereas the response
to a perturbation which acts on infinite timescales (i.e. diverging as $t,t'$) is finite.
More precisely: 
\begin{equation}\label{wltm}
\lim_{t\rightarrow \infty }\int_{0}^{t^{*}}R(t,u)f (u)du=0 \qquad {\mbox {but}}
\qquad \lim_{t\rightarrow \infty }\int_{0}^{t}R(t,u)f(u)du\neq 0 \quad ,
\end{equation}
where $f(t)$ is a generic function. Therefore the dynamics on
``infinite timescales'' ($t,t'\rightarrow \infty $) decouples from the
transitory regime.\\ 
Finally, we note that more general Ans{\"a}tze with a set of different diverging timescales have been 
used in the context of classical \cite{CugliandoloKurchanPhilMag,FranzMezard} and quantum 
\cite{ChamonRotors} glassy systems. However, this type of solutions are physically and
technically related to a full replica symmetry breaking solution in the thermodynamical 
analysis. For systems characterized by a one step replica symmetry breaking solution
in the thermodynamics, as the model we are focusing on \cite{SGletter}, 
one generally expects only one diverging timescale.

\subsubsection{Generalized QFDR}\label{GQFDR}
An outstanding physical property of the asymptotic dynamical solutions, 
discovered in the classical case by Cugliandolo and Kurchan
\cite{CugliandoloKurchanPRL} and in the quantum case by Cugliandolo and
Lozano \cite{CugliandoloLozano}, is that in the aging time-sector
($t$, $t'$ and $t-t'$ are large 
but the ratio $h(t')/h(t)$ stays finite) the standard fluctuation dissipation
relation is violated but there exists a generalized fluctuation dissipation relation
(GFDR) between the 
correlation and the response functions which has
the usual functional form of the FDR (generalized
to non TTI functions) and in which the 
temperature $T$ is replaced with an effective temperature $T_{\text{eff}}$
\cite{CugliandoloLozano}.
Two remarks are in order concerning $T_{\text{eff}}$. First, a physical one:
the effective temperature has a real physical meaning of temperature
\cite{effectivetemp} since is what a thermometer, whose reaction time
equals the  
timescales on which the aging evolution takes place, would measure.
Second, a technical one. The effective temperature $T_{\text{eff}}$ is related to 
the {breaking point $x$} arising in the replica symmetry breaking solution 
of the thermodynamics, i.e. $T_{\text{eff}}=T/x$.
A general argument to show why one expects this 
to be true for a very large class of classical 
systems, included finite dimensional
systems, has been presented in \cite{FranzMezardParisiPeliti}. \\
It is important to note, as pointed out in \cite{CugliandoloLozano}
that in the quantum case the GFDR is expected to become classical. The
argument  
is the following: if 
$T_{\text{eff}}$ is finite then the Fourier integral relating the correlation and 
the response is dominated by $\omega \propto 0$. Hence, one can develops 
the hyperbolic tangent recovering back a classical generalized fluctuation
dissipation relation, which in our case (for spins) reads:
\begin{equation}\label{gcfdr}
R_{S}(t,t')=\frac{1}{T_{\text{eff}}}\partial _{t'}C_{S}(t,t')
\end{equation} 
This becomes a relation between the aging functions:
\begin{equation}\label{gfdr2}
R^{AG}_{S}(\mu )=\frac{1}{T_{\text{eff}}}(C^{AG}_{S})'(\mu )
\end{equation}
where $\mu =h(t')/h(t)$.
Moreover this has been argued to be generically true for models with simple
commutation relations (particles and rotors) in \cite{ChamonRotors} since 
the dynamical equations are fixed point of the re-parameterization group
of time transformations and the {\it renormalized} aging dynamics becomes 
classical at the fixed point. The quantum mechanics enters
only as a renormalization of the coefficients of the dynamical equations.
We will show that this is also the case for our system which is characterized
by non trivial commutation relations between the spins (contrary to the case
of rotors or particles). However, the behavior next to the quantum
critical point is still unclear (See section \ref{Conclusion}).

\subsection{Numerical results}
\begin{figure}[ht]
\[
\figx{14cm}{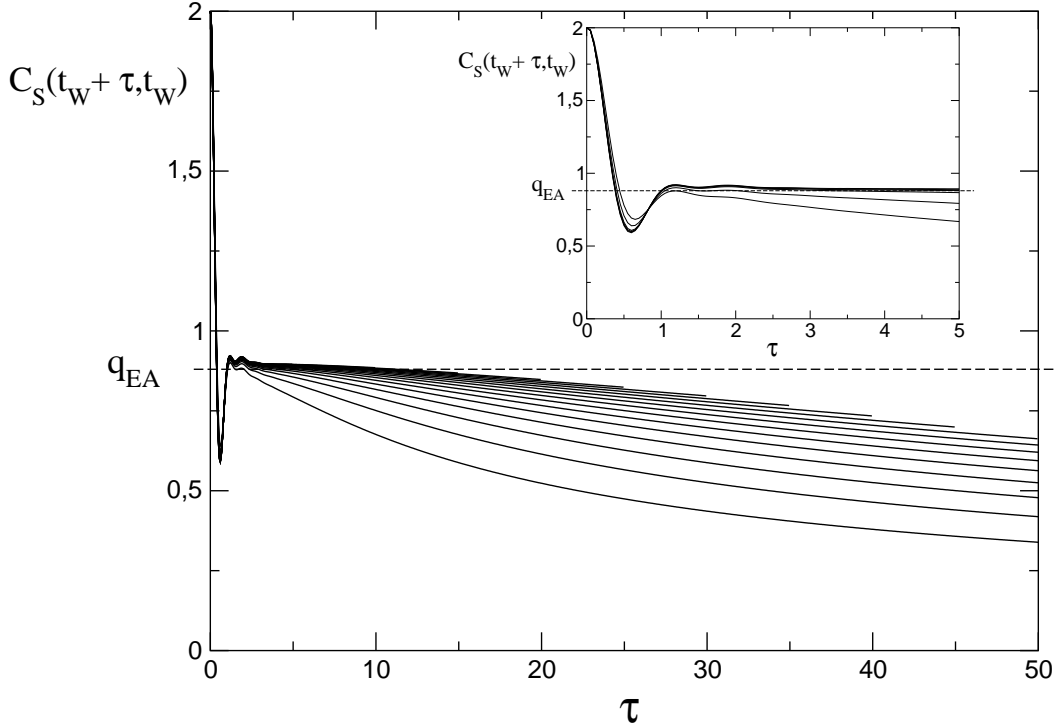}
\]
\caption{\label{CAG.fig}
\sl Spin correlation as a function of $\tau $ for $t_{w}=15,20,\dots
,95$ (from bottom to top ($t_{max}=100$)), $J_{B}=J_{H}=S=1$ and
$T=0.1$ (inside the glassy phase).  
The height of the dotted straight line
equals the Edwards-Anderson parameter computed within the static formalism and
coincides well with the plateau value. The aging behavior is explicit.  
Inset: zoom of the spin correlation as a function of $\tau $ 
on the time interval corresponding to the stationary regime for
$t_{w}=5,10,25,37.5,50,75$ from bottom to top. The curves show a 
 clear convergence toward a TTI stationary regime.
}
\end{figure}
The numerical procedure used is described in Appendix
\ref{AppendiceNumerique}. It turns out that the problem is more
difficult to solve than the classical ones or the quantum
$p-$spins model, because the dynamical equations are for the auxiliary
boson $b$ and not directly for the physical spin $S$. 
\begin{figure}[ht]
\[
\figx{14cm}{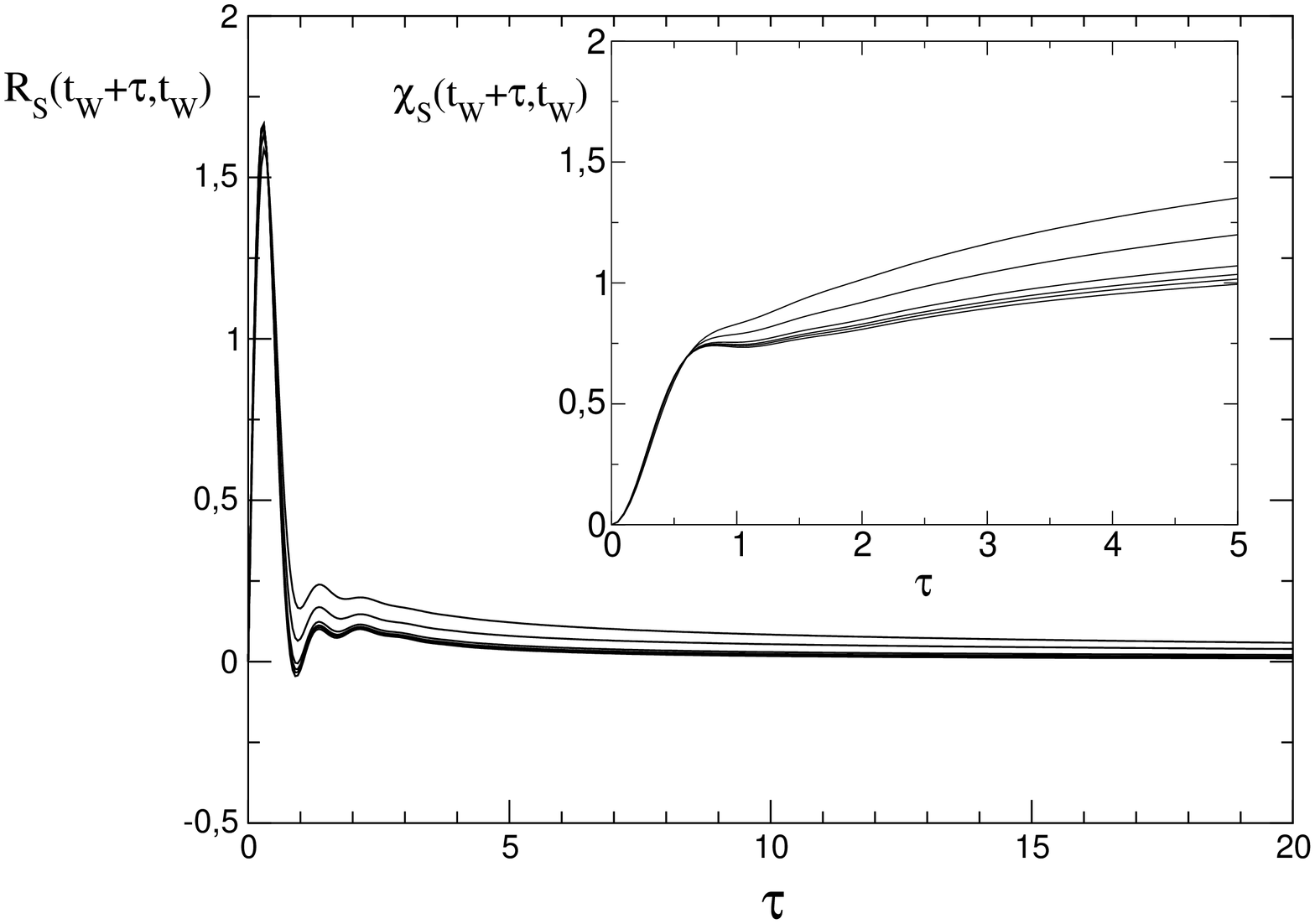}
\]
\caption{\label{RAG.fig}
\sl Spin response as a function of $\tau $ for  $t_{w}=5,10,25,37.5,50,75$ 
(from top to bottom ($t_{max}=100$)), $J_{B}=J_{H}=S=1$ and $T=0.1$ (inside the glassy phase). 
Inset: zoom of the integrated response as a function of $\tau $ 
on the time interval corresponding to the stationary regime for
$t_{w}=5,10,25,37.5,50,75$ from bottom to top. The aging behavior and
the weak long term 
memory scenario are explicit.
}
\end{figure}
As clearly indicated
in the Ansatz, even in the aging time-regime where the spin correlation
and response function evolve very slowly, the bosonic functions oscillate 
wildly. 
\begin{figure}[ht]
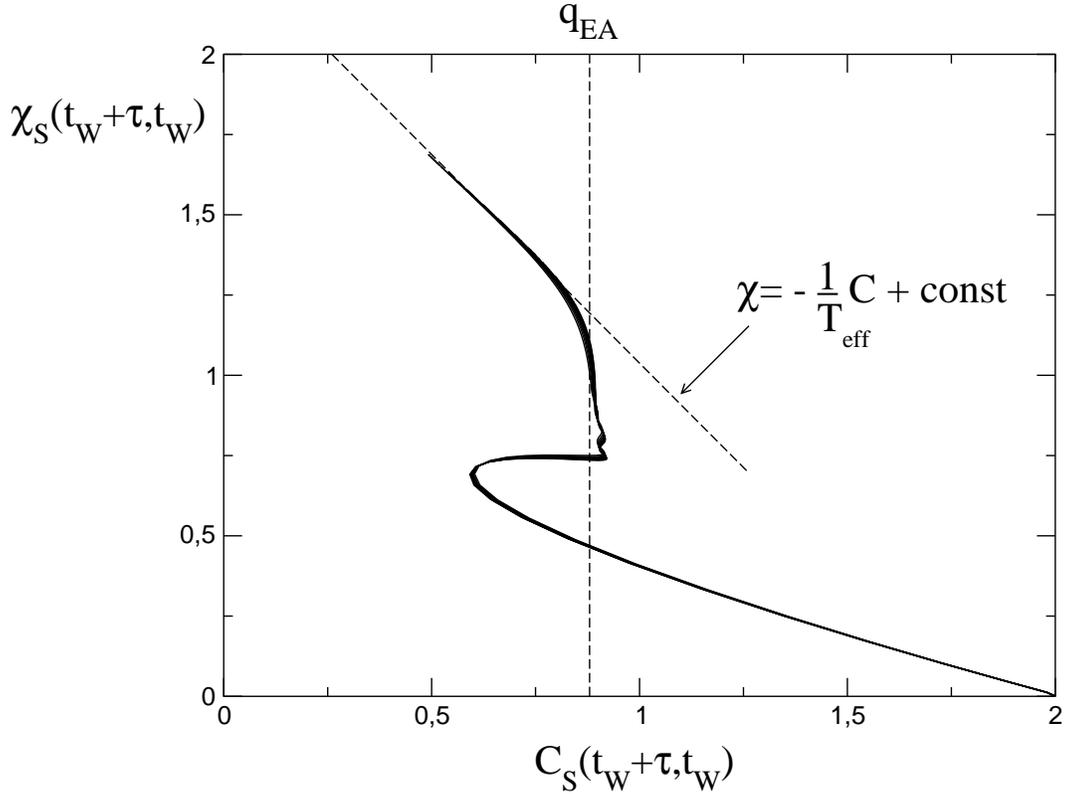

\[
\figx{14cm}{ChiVsC.eps}
\]
\caption{\label{Chi.fig}
\sl Parametric plot of the spin integrated response as a function of $C_{S}$
for  $t_{w}=40,45,55,60,65,70$ ($t_{max}=100$), $J_{B}=J_{H}=S=1$ and
$T=0.1$ (inside the glassy phase). The collapse predicted  
by the dynamical Ansatz is good. The vertical dotted straight line  
indicates the values of the Edwards-Anderson computed within the static
formalism. The dashed straight line has a slope $-1/T_{\text{eff}}$ where 
$1/T_{\text{eff}}=x/T$ and $x$ has been computed within the static
formalism. The curves clearly show that the generalization of 
the fluctuation dissipation relation holds in the aging regime.
}
\end{figure}
The numerical solution thus demands a more sophisticated algorithm, 
inspired from well-known methods to solve one variable differential equations.\\
The numerical results support and validate completely the out of equilibrium
scenario encoded in the Ansatz (\ref{Ansatz}). In the following we present 
the results for $J_{H}=1$, $J_{B}=1$ and a temperature well inside in the
glassy phase $T=0.1$. 
In Fig. \ref{CAG.fig},
we plot respectively the spin correlation function 
$C_{S}(t_{w}+\tau ,t_{w})$ and the spin integrated response function 
$\chi(t_{w}+\tau ,t_{w})=\int_{t_{w}}^{t_{w}+\tau }ds R_{S}(t_{w}+\tau ,s)$ as 
a function of $\tau $ for different values
of $t_{w}$. We remark that a pseudo stationary regime sets in for
$\tau <  1$ with a plateau, whose height is the Edwards-Anderson
parameter. Its value ($q_{EA}^{st}\simeq 0.88$), computed from the static
analysis by the marginality prescription, is represented with a
dashed line on Figure \ref{CAG.fig} : this shows a very good agreement
between the two methods.
\begin{figure}[ht]
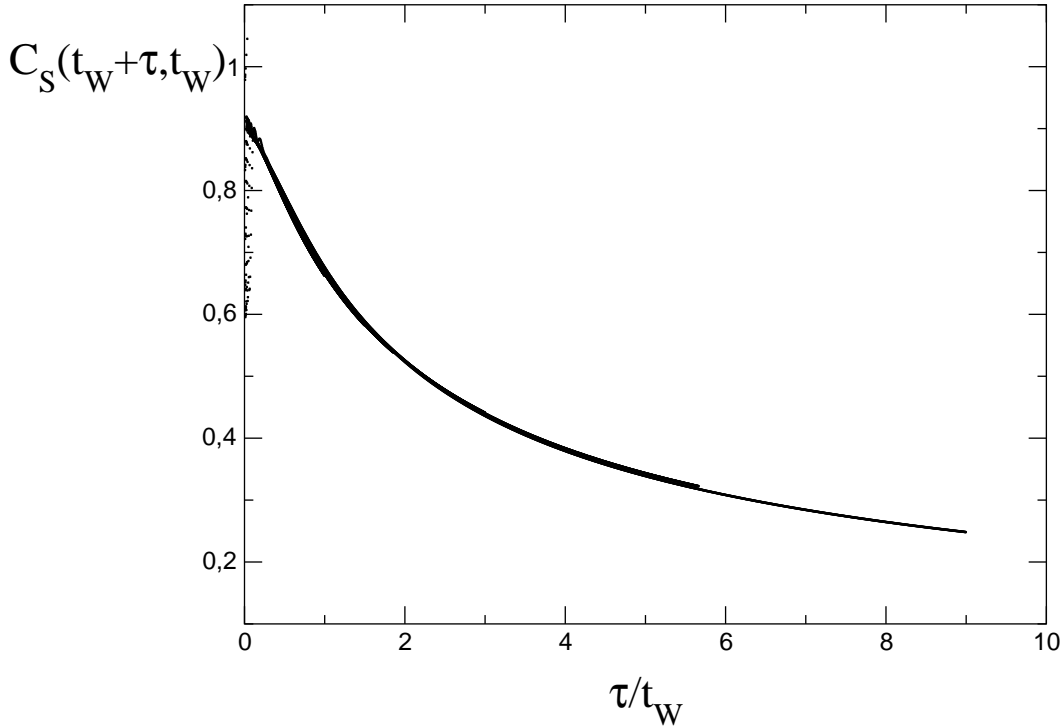

\[
\figx{14cm}{Scaling.eps}
\]
\caption{\label{Scaling.fig}
\sl Spin correlation as a function of $\tau/t_{w}$ for
$t_{w}=15,20,\dots ,95$ (from bottom to top ($t_{max}=100$)),
$J_{B}=J_{H}=S=1$ and $T=0.1$. These 
are the same curves plotted in Fig. \ref{CAG.fig} but with respect to the
variable $\tau/t_{w}$. The excellent collapse strongly suggests that the
function $h(t)$ (present in the dynamical Ansatz) should be equal to $t$. 
}
\end{figure}

\begin{figure}[ht]
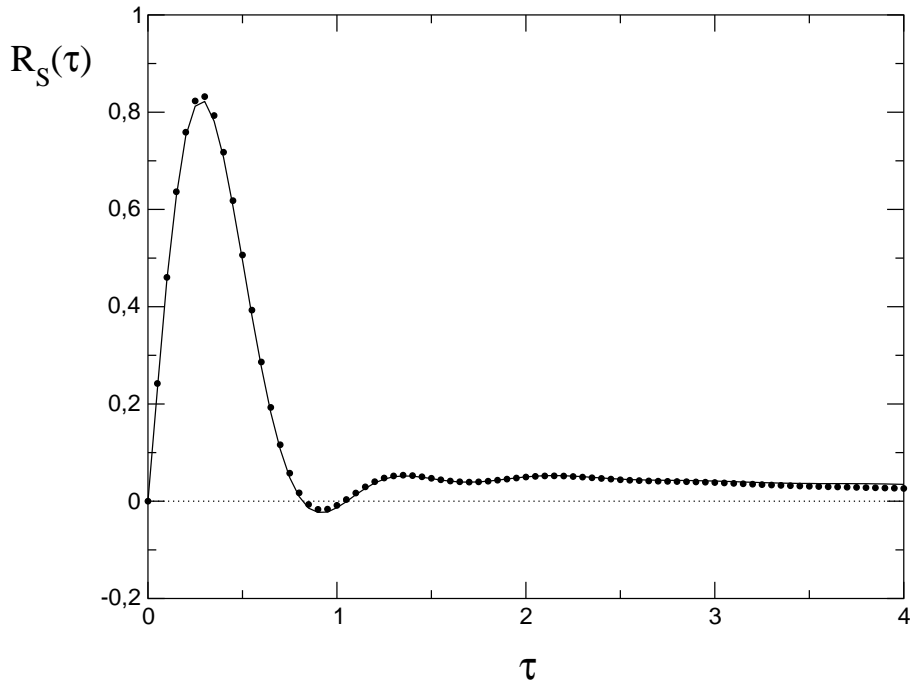

\[
\figx{12cm}{qfdt_tti.eps}
\]
\caption{\label{qdft_tti.fig}
\sl
TTI part of the 
spin response function (dots) compared to 
its expression (continuous line) computed by the QFDR from the correlation for
$J_{B}=J_{H}=1$, $S=1$, $t_{W}>10$ and $T=0.1$. 
}
\end{figure}

Moreover the fact that correlation and
response are related by the QFDR in this time sector, see
Fig. \ref{qdft_tti.fig}, 
shows nicely that this is indeed a pseudo-equilibrium regime.

On a longer timescale $\tau \propto t_{w}$
the aging behavior sets in as clearly indicated in the two figures. 
The integrated response and the correlation evolve more and more slowly
increasing $t_{w}$. Note that the behavior of $\chi$  shows explicitly
the weak long term memory scenario: even if the response vanish
in the aging time sector the integrated response does not. Therefore
the magnetization response to a constant and small magnetic field 
switched on at
$t_{w}$ depends always explicitly on $t_{w}$ and becomes slower
 increasing $t_{w}$.
Moreover we note that the numerical results strongly 
suggest that $h(t)=t$. Indeed the different curves $C_{S}(t_{w}+\tau,t_{w})$
collapse very well on a single curve ($K_{AG}(\mu )$) when plotted as a 
function of $(t_{w}+\tau) /t_{w}$, see Fig. \ref{Scaling.fig}. We have also verified the
 correctness of the GFDR hypothesis. In Fig. \ref{Chi.fig}, we show a parametric
 plot of $\chi$ as a function of $C_{S}$ for different values of $t_{w}$.
For classical systems \cite{reviewDYN} the limiting curve (for
$t_{w}\rightarrow \infty $) is very useful to characterize the aging
behavior. For systems with only one diverging timescale, because of the form
of the FDR and the GFDR, one finds a two straight line
plot where $\chi\propto -\frac{1}{T}C_{S}$ for $q_{EA}<C_{S}<C(t,t)$ and
$\chi\propto -\frac{1}{T_{eff}}C_{S}$ for $0<C_{S}<q_{EA}$. In the quantum case the situation is more
involved since in the stationary regime $R_{S}$ and $C_{S}$ are related
by the QFDR, therefore one does not expect that for $q_{EA}<C_{S}<K(t,t)$
the plot should be very useful. 
However, as discussed in
\cite{CugliandoloLozano}, in the aging time sector where
$0<C_{S}<q_{EA}$ the GFDR becomes classical 
and one should recover a straight line for $\chi$ whose slope equals
the inverse of the effective temperature. This is indeed what we find in 
Fig. \ref{Chi.fig}, where we compare the behavior of $\chi$ at low $C_{S}$ with 
a straight line whose slope is $1/T_{\text{eff}}=x/T$. 
\begin{figure}[ht]
\[
\figx{14cm}{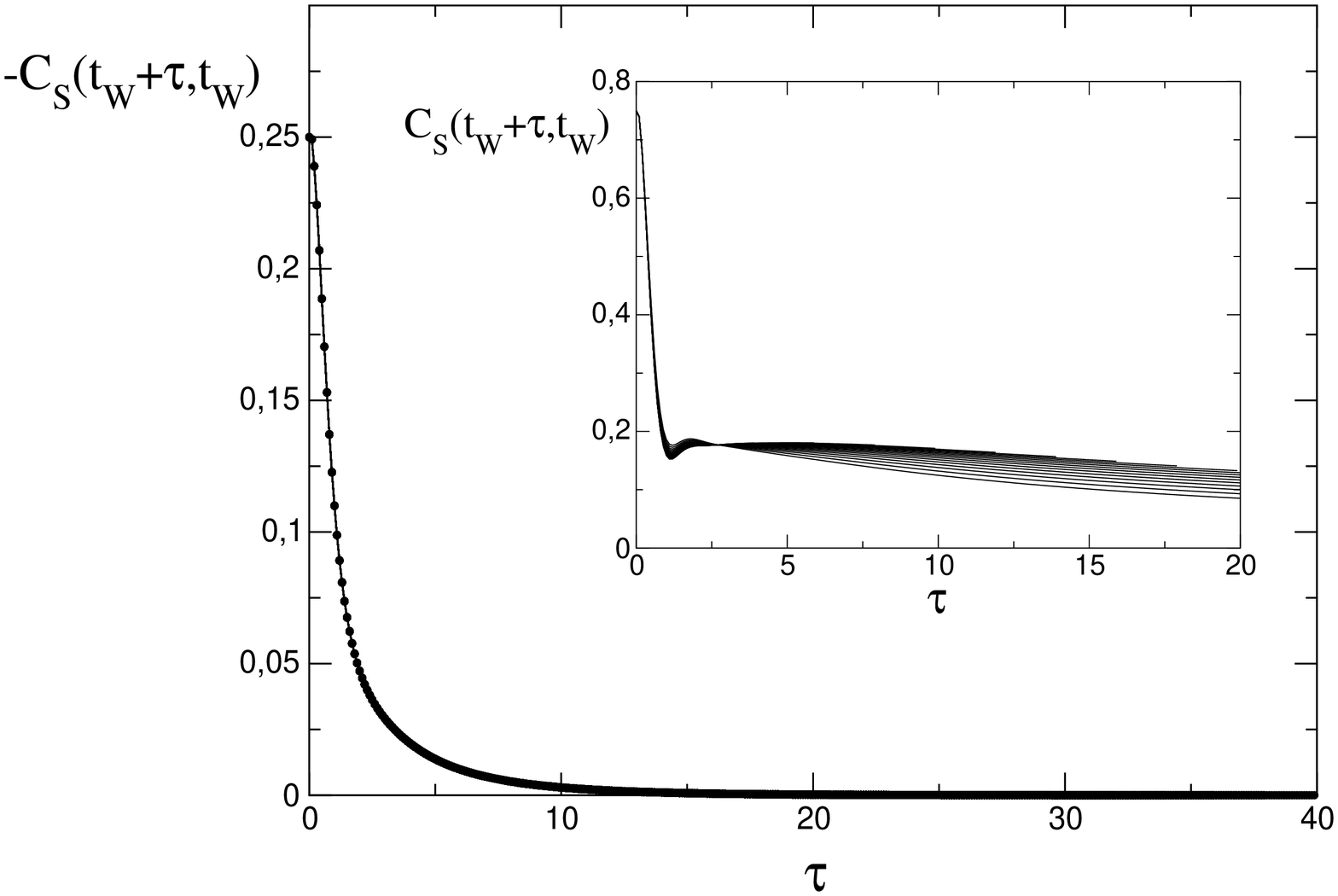}
\]
\caption{\label{Fermions.fig}
\sl Spin correlation as a function of $\tau $ in the fermionic case
for  $t_{w}=12,14,\dots 48$ ($t_{max}=100$), $J_{B}=J_{H}=1$, $S=0.5$ and
 $T=0.1$. Inset: Spin correlation as a function of $\tau $ in the bosonic
case for the same value of the parameter. This figure clearly shows 
the existence of aging in the bosonic case and 
its absence in the fermionic one.}
\end{figure}
$x$ is the breakpoint in 
the one-step replica symmetry breaking solution obtained (for the same
value of the parameters) generalizing 
 the analysis performed in \cite{SGletter} to taking into account the 
presence of the bath  (see Appendix \ref{AppendiceMatsubara}).

Finally, we can obtain the  equations for the fermionic model by the
simple change $S\rightarrow -S, J_{H}^{2}\rightarrow  -J_{H}^{2}$  in
(\ref{EqsGenerales})
as discussed previously. Numerically solving these equations we have found no 
glassy behavior as predicted in \cite{SachdevYe}. Indeed
we show in Fig. \ref{Fermions.fig} the spin correlation function in the fermionic case, which
does not show any aging behavior at low temperature, whereas the same calculation for bosons (with the
same parameters) clearly does.

\subsection{Analysis of the stationary regime}\label{stationary}
We focus now on the time sector in which the difference between $t$ and $t'$
stays finite and $t,t'$ are very large. Hence, in
Eqs. (\ref{ansatz1},\ref{ansatz2}) the aging part does not evolve and
is zero for the response and  
equals $-ig$ for the correlation. Plugging the Ansatz 
(\ref{ansatz1},\ref{ansatz2}) into the dynamical equations we get:
\begin{eqnarray}\label{stationaryeq1}
(i\partial +\lambda )R_{ST}(t)&=&\int_{0}^{t}dt'\Sigma _{R}^{ST}(t-t')R_{ST}(t')
+\delta (t)\\
(i\partial +\lambda )K_{ST}(t)&=&\int_{-\infty }^{+\infty }dt'\Sigma
_{R}^{ST}(t-t')K_{ST}(t')+\int_{-\infty }^{+\infty }dt'\Sigma
_{K}^{AG}(t+t')R_{ST}^{*}(t')+A\label{stationaryeq2}\\ 
A&=&\lim_{t\rightarrow \infty }\int_{0}^{t}du\frac{h'(u)}{h(t)}\Sigma
_{R}^{AG}\left(\frac{h(u)}{h(t)} \right)K_{AG}\left(\frac{h(t)}{h(u)}\right)+\lim_{t\rightarrow \infty
}\int_{0}^{t}du\Sigma _{K}^{AG}\left(\frac{h(u)}{h(t)} \right)\frac{h'(u)}{h(t)}R_{AG}^{*}\left(\frac{h(u)}{h(t)} \right)\nonumber\\ 
&&-ig\int_{0}^{\infty }dt'\Sigma _{R}^{ST}(t')+\Sigma _{K}^{\infty
}\int_{0}^{\infty }dt'R^{ST}(t')+i\lambda g\label{stationaryeq3} 
\end{eqnarray} 
where we have used for the self-energy the same notation introduced in
(\ref{ansatz1},\ref{ansatz2}), and $\Sigma _{K}^{\infty }=\Sigma _{K}^{AG}(1)$.
Since we have defined $K_{ST}(t)$ in such a way 
that it vanishes in the long time
limit, $A$ has to be equal to zero. Note that this overall equation
 couples the stationary and the aging regime. 
As in the paramagnetic case, one can show (order by order in
perturbation theory) that $R_{ST}$ and $K_{ST}$ 
satisfy the QFDR (and therefore $R_{S}^{ST},C_{S}^{ST}$ too). It could
seem that the procedure to fix $\lambda $ is different in the two
cases. In fact in 
the paramagnetic case one chooses $ \lambda $ in such a way that the 
QFDR is verified for bosonic functions, instead now $\lambda $ is 
such that the correlation and response functions do not oscillate
 in the large time limit when $t-t'$ and $t,t'$ are very large. 
But the two values of $\lambda$ are the same an asymptotically oscillating
function cannot satisfy the QFDR relation.

As a conclusion the stationary equations can be fully interpreted 
as equilibrium dynamical equations. Indeed it has been shown in the classical 
\cite{CugliandoloKurchanPRL} and recently in the quantum case \cite{QuantumTAP}
that this type of equations represents the pseudo-equilibrium relaxation inside
the marginally stable TAP states (local minima of the free energy landscape 
whose Hessian is characterized by a vanishing fraction of zero modes).
Imposing the marginality condition in the static computation \cite{SGlong}
is equivalent to consider a Boltzmann measure restricted to the marginally
stable TAP states.  It is for this reason that one can 
get information about the out of equilibrium dynamics 
by a purely equilibrium computation.
However, it is important to understand that the equations
(\ref{stationaryeq1},\ref{stationaryeq2},\ref{stationaryeq3}) do not really
represent an equilibrium relaxation. Because the marginally stable TAP states 
have a vanishing fraction of zero modes, the system
find always a way to ``escape'' to these states,
even if more and more slowly and this gives rise to the aging behavior. 
Hence, the physical mechanism inducing the slow dynamics 
is not an activated jump dynamics
across some energy barriers but it is an entropic effect.
The slow dynamics of the system
is due to the fact that the longer is the time the smaller is 
the number of directions along which the system can escape.
Finally, the fact that the marginally stable TAP states dominate the 
off-equilibrium dynamics whereas they are not relevant for equilibrium 
properties helps to understand why the dynamical transition temperature 
is different from (actually is larger than) the equilibrium transition 
temperature. 
In fact after a quench, the system is almost trapped
in those minima and thus displays the aging phenomenon at long time. 
The local minima responsible for the slow dynamics
appear at a temperature higher than $T_{eq}$, thus $T_{d}> T_{eq}$.
The activated dynamics which would probably restore
the equality $T_{d}=T_{eq}$ is on timescales diverging with $N$, 
completely unaccessible to our 
mean-field analysis.

\subsection{Analysis of the aging regime}\label{aging}
Let us now focus on the aging regime, i.e. $t$, $t'$ and also $t-t'$ are large
but the ratio $h(t')/h(t)$ stays finite. Note that 
within the following asymptotic analysis
one cannot find out what is the function $h(t)$
\cite{CugliandoloKurchanPRL,CugliandoloLozano,ChamonRotors}. This is indeed an
open problem already for 
classical systems. However, the numerical results, see
FIG. \ref{Scaling.fig}, suggest
that for our model $h(t)=t$.

Plugging the Ansatz (\ref{ansatz1},\ref{ansatz2}) into the dynamical equations
and after some manipulations similar to 
 \cite{CugliandoloLozano} we get:
\begin{align}\label{agingeqs1}
\lambda R_{AG}(\mu )=&\int_{\mu }^{1}\frac{d x}{x}\Sigma
_{R}^{AG}(x)R_{AG}\left( \frac{\mu }{x} \right)+\Sigma
_{R}^{ST}(\omega =0)R_{AG}(\mu )+\Sigma _{R}^{AG}(\mu ) 
R_{ST}(\omega =0)\\
\lambda K_{AG}(\mu )=&
\int_{0}^{\mu }\frac{d x}{\mu}\Sigma _{K}^{AG}(x )
R_{AG}^{*}\left( \frac{x}{\mu }\right)
+\int_{0}^{1}dx
\Sigma _{R}^{AG}(x)K_{AG}\left(\frac{\mu }{x}\right)
+\Sigma _{R}^{ST}(\omega =0)K_{AG}(\mu )\nonumber
\\
&+\Sigma _{K}^{AG}(\mu )
R_{ST}^{*}(\omega =0)\label{agingeqs2}\\
\Sigma _{R}^{AG}(\mu )=& - \frac{J_{H}^{2}}{4}\left(K_{AG}^{2}(\mu )R_{AG}^{*}
(\mu )-2 |K_{AG}(\mu )|^{2}R_{AG}(\mu )\right)\label{agingeqs3}\\
\Sigma _{K}^{AG}(\mu )=& \frac{J_{H}^{2}}{4}|K_{AG}(\mu )|^{2}K_{AG}(\mu )
\label{agingeqs4}
\end{align}
where $K_{AG}$ satisfies the boundary condition $K_{AG}(1)=-ig$ and we
have used the notation $\mu =h(t')/h(t)$. It is important to remark
that: 
\begin{enumerate}
\item  there is 
no bath contribution to the aging part of the self-energy. This is 
natural and it is probably generally true since the bath has always its own
equilibration timescale, therefore in the aging time-sector ($t$, $t'$ and
 $t-t'$ are large but the ratio $h(t')/h(t)$ stays finite) the bath 
is always already equilibrated and cannot give a non-constant aging
contribution.
\item  the terms
linear in $R$ (and higher) have been neglected in $\Sigma _{K}^{AG}$, whereas
the terms quadratic (and higher) have been neglected in $\Sigma _{R}^{AG}$. 
This is due to the fact that they do not give a finite contribution in the 
aging equations.
\item As pointed out in the classical case \cite{CugliandoloKurchanPRL}
and recently in the quantum case \cite{ChamonRotors} the equations
(\ref{agingeqs1},\ref{agingeqs2},\ref{agingeqs3},\ref{agingeqs4}) are
re-parameterization invariant.  However there
is only one function $h(t)$ reached by the system in the long time limit. 
This is a general problem arising in the study of the asymptotic solution of 
partial and integro-differential equations, called the matching problem. Until 
now different techniques are known and applied to solve this problem 
for partial differential equations 
 but its solution for the dynamical equations arising 
in the study of glassy systems remains an open problem. 
\item The correlation and response functions are supposed to be related by
the GFDR in the aging regime 
after that the exponential $e^{-i\lambda (t-t')}$ has been gauged
out. In particular the GFDR predicts that $R_{AG}(\mu
)=-\frac{i}{2T_{\text{eff}}}(K_{AG})'(\mu )$ (in our notation the GFDR for
bosons has a $-i/(2T_{\text{eff}})$  
instead that $1/T_{\text{eff}}$). One can indeed verify that this is really 
a property of the aging equations: (\ref{agingeqs1}) can be obtained 
by differentiating (\ref{agingeqs2}) and using the GFDR. Moreover, we remark 
 that if the bosonic correlation and response
functions verify the GFDR so do the spin correlation and response
functions as in (\ref{gfdr2}) : $T_{\text{eff}}$ is the effective temperature.
\end{enumerate}
Evaluating the aging equations in $\mu =1$ and imposing the existence of 
an aging solution, i.e. $R_{AG}(1)\neq 0$ we obtain two matching conditions
with the stationary regime \cite{CugliandoloLozano}: the first one is 
\begin{equation}\label{aingeqs5}
\lambda =\Sigma _{R}^{ST}(\omega =0)+\left. \frac{\Sigma _{R}^{AG}}{R_{AG}}
\right|_{\mu =1}R_{ST}(\omega =0)
\end{equation}
and the second is the same one already obtained from the 
long-time limit of the stationary equation, i.e. $A=0$, (eq. (\ref{stationaryeq3})).
One can simplify further these two matching equations. Indeed,
thanks to the GFDR (\ref{gfdr2}), one can perform the integrals on the aging 
functions in $A$. Hence, using the zero frequency term of (\ref{stationaryeq1}), 
we get:
\begin{eqnarray}\label{match2a}
\frac{J_{H}^{2}}{4}R_{ST}^{2}(\omega
=0)g^{2}\left(\left.\frac{R_{AG}^{*}}{R_{AG}} \right|_{\mu
=1}+2\right)&=&1\\ 
\frac{i}{2T_{\text{eff}}}\Sigma _{K}^{\infty }g-\frac{ig}{R_{ST}(\omega =0)}-
\Sigma _{K}^{\infty }R_{ST}^{*}(\omega =0)&=&0
\label{match2b} 
\end{eqnarray}
Moreover, because
 $R_{ST}$ is a bosonic response function in a pseudo-equilibrium 
regime, its zero frequency component is real and negative. Therefore
$\left.\frac{R_{AG}^{*}}{R_{AG}} \right|_{\mu =1}$ is a real number equal to one 
or minus one that we will 
note $\zeta  $ in the following and $R_{ST}(\omega =0)$ reads:
\begin{equation}\label{rst2}
R_{ST}(\omega =0)=-\frac{2}{J_{H}g\sqrt{2+\zeta }}
\end{equation}
Plugging this expression onto (\ref{match2b}) and using that 
$\Sigma _{K}^{\infty}=-i\frac{J^{2}_{H}}{4}g^{3}$ we finally obtain the equation
for $T_{\text{eff}}$. The aging solution corresponds to $\zeta =+1$ ($\zeta =-1$
implies $g^{2}/T_{\text{eff}}=0$) and is characterized by the following equation:
\begin{equation}\label{qea}
\frac{J_{H}q_{EA}}{T_{\text{eff}}}=\left(\sqrt{3}-\frac{1}{\sqrt{3}} \right)
\end{equation} 
where we have replaced $g^{2}=4q_{EA}$. Note that, replacing
$T_{\text{eff}}$ with $T/x$ (where $x$ is the breakpoint in the one-step
replica symmetry breaking scheme) this becomes the same equation
obtained in \cite{SGletter} using the marginality condition.\\
Indeed  (\ref{stationaryeq1},\ref{stationaryeq2}) with $A=0$ and
(\ref{rst2},\ref{qea}) and the boundary condition $K_{ST}(\tau =0)=-i(2S+1)+ig$
are a closed set of equations that completely determines 
$K_{ST},R_{SR},\lambda ,T_{\text{eff}},q_{EA}$.  In Appendix
\ref{AppendiceMatsubara}, we show that 
they are completely equivalent to the equations studied in
\cite{SGletter,SGlong} using the marginality prescription within  
a pure static computation.
Finally, let us stress that, even if $J_{B}$ is not present in eq. (\ref{qea}),
$T_{\text{eff}}$ depends on $J_{B}$ because the Edwards-Anderson parameter 
depends on $J_{B}$ via the eqs. (\ref{stationaryeq1},\ref{stationaryeq2}) 
which contains explicitly this coupling constant.

\section{Role of the relaxation bath}\label{RoleOfTheBath}

In this section, we  briefly discuss  the effect of the coupling
strength to the relaxation bath $J_{B}$. 
In deriving Eqs (\ref{EqsGenerales}), we took a generic bath and
expand in second order in its coupling constant $J_{B}$. Thus  the
equations we derived are a priori only valid in the limit of small
$J_{B}$.
However, we will show that our main equations (\ref{EqsGenerales})
also describe the dynamics of a model with finite $J_{B}$ in a extreme
limit. Thus it is legitimate to study them for $J_{B}$ finite (there
is no risk of inconsistencies).

The effect of the bath will of course depend on its
precise form. Two types of bath can be considered :  baths that
only couple to the spin, and baths that couple to the boson or
the fermions. In this discussion, we will concentrate on the first
kind, since the spin is the physical object, not the boson.
 One of the simplest possibility is to couple the
spin to 2 fermions using the Kondo interaction. In order to take the
large-$N$ limit, we directly introduce the $SU(N )\times SU (N\gamma )$
Kondo model, with  $N\gamma$ flavors, defined by
\cite{ParcolletGeorgesPRLKondo+ParcolletGeorgesKotliarSengupta}  : 
\begin{equation}\label{DefBathKondo}
H = 
\frac{1}{\sqrt{{\cal N}N}} \sum_{i<j} J_{ij} \vec{S}_{i}\cdot
\vec{S}_{j} 
+
 \sum_{
\substack{k\\
1\leq  i \leq N\gamma \\
1\leq  \alpha\leq N}}
 \epsilon_{k}  c^{\dagger}_{ki\alpha }c_{ki \alpha  }
+
\frac{J_{B}}{N\sqrt{\gamma }}\sum_{\substack{1\leq\alpha,\beta \leq N\\ 1\leq  i \leq
N\gamma}} S_{i\alpha \beta } c^{\dagger}_{ki\beta }c_{k'i\alpha }
\end{equation}
where $c$ are the bath fermions, $\epsilon_{k}$ their kinetic energy and
$J_{B}$ is now the Kondo coupling. 
In the large-$N$ limit, one can still find a closed system of
equations, but at the expense of introducing  auxiliary fermionic
Green functions $r$ and $k$, as explained for example in  \cite{ParcolletGeorgesPRLKondo+ParcolletGeorgesKotliarSengupta}.
The dynamical equations are similar to (\ref{EqsGenerales}) :
(\ref{DysonR},\ref{DysonK},\ref{Boundary}) are the same  and the bath
term in self energies
(\ref{SigmaLORb},\ref{SigmaLOKb}) are replaced by : 
\GroupeEquations{
\begin{align}\label{SigmaLOKondo}
\Sigma_{R}  &= 
- \frac{J_{H}^{2}}{4}
\biggl(
     \Bigl ( K^{2} + R^{2}\Bigr ) R^{*} - 2 |K|^{2} R
\biggr)
- \frac{i\sqrt{\gamma} J_{B}}{2}
\biggl(
    R_{0}^{*} k - K_{0}^{*} r 
\biggr)
\\
\Sigma_{K}  &= 
- \frac{J_{H}^{2}}{4}
\biggl(
      2 K |R|^{2} - K^{*}
      \Bigl (R^{2} + K^{2} \Bigr )
\biggr)
  - \frac{i\sqrt{\gamma} J_{B}}{2} 
\biggl(
   R_{0}^{*} r     -  K_{0}^{*} k 
\biggr)
\end{align}
}
Since the bath has now a proper dynamics, $r$ and $k$ should be computed  using
new Dyson equations : 
\GroupeEquations{
\begin{align}\label{DysonAuxKondo}
r (t,t') &= \delta (t-t') + \int_{t'}^{t} du \; \sigma_{R} (t,u)r(u,t')
\\
k (t,t') &= 
\int_{0}^{t} du \,\,  \sigma_{R} (t,u)k (u,t')
+
\int_{0}^{t'} du \,\, 
\sigma_{K} (t,u) r(t',u)^{*}
\end{align}
}
and their self-energy reads: 
\GroupeEquations{
\begin{align}\label{EquationsLOBasis}
\sigma_{R}^{\text{Heis}}  &= 
 \frac{iJ_{B}}{2\sqrt{\gamma}}
\biggl(
    R_{0}K  + K_{0} R 
\biggr)
\\
\sigma_{K}^{\text{Heis}} &= 
 \frac{iJ_{B}}{2\sqrt{\gamma}} 
\biggl(
    R_{0 } R + K_{0} K 
\biggr)
\end{align}
}
$r$ should also satisfy the boundary condition :
\begin{equation}\label{Boundaryr}
\lim_{t\rightarrow (t')^{+}}r (t,t')= -i
\end{equation}

It is not difficult to show that in the limit of an infinite number
of channel $\gamma \rightarrow \infty$, these equations reduce to
(\ref{EqsGenerales}).

However, for finite $\gamma$, these equations are much more complex than
(\ref{EqsGenerales}), since at low temperature the Kondo scale appears
and one has to deal with a problem with many different scales. This
really increases the difficulty of a numerical computation.
However, one can extend rather simply 
the previous analytical study and verify that the same
dynamical scenario continues to hold.
In this paper, we restricted ourselves to study
(\ref{EqsGenerales}) as a function of the strength
of the bath $J_{B}$, using Matsubara formalism with marginality
condition \cite{SGlong}. We found that increasing $J_{B}$ the
spin glass transition temperature $T_{g}$ decreases 
monotonically and as far as we can
solve the numerical equations, the transition is still second order, given by the condition
$x=1$ ($x$ is the value of the breakpoint in the replica
formalism). However, the decrease of $T_{g}$ is slow and we could not
reach numerically a point where  it vanishes. Numerical computation
can not for the moment decide whether there is a second order phase
transition until a quantum critical point at finite $J_{B}$ and $T=0$ or the
spin glass is not destroyed at zero temperature until $J_{B}=\infty$.
As emphasized in our concluding remarks, this situation is
disappointing since we would like to study the aging in  the  vicinity of a non
pathological quantum critical point for this model.
Our interpretation is that this bath, coupled to the spin directly, is
not ``efficient'' enough and that we probably need to couple to a bath
with charge fluctuations by introducing holes in the model. Such a
doped model is also interesting physically to study the destruction of
a quantum spin glass by doping. 
Finally, let us emphasize that solving numerically the model with the
Kondo bath (\ref{DefBathKondo}) or a more general bath may lead to
more interesting results, such as an increase of the critical
temperature and the Edwards-Anderson parameter as the coupling to
the bath increases from 0 \cite{GrempelCugliPRIVATE}.

\section{Summary and discussion}\label{Conclusion}
In this paper we have studied the out of equilibrium dynamics of the quantum Heisenberg spin glass 
defined on a completely connected lattice and coupled to a spin thermal bath.  
We have replaced the $SU (2)$ spin symmetry group with $SU (N)$ and we have considered the large-$N$ limit. 
This has allowed us to have a more tractable model which however seems
to capture some of the physics of
 the $SU (2)$ case \cite{SGlong}. Thanks to the large-$N$ limit we have obtained a set 
of closed integro-differential equations on the correlation and
response functions. By the analytical  
study and the numerical integration of these equations we have fully analyzed the 
real time (dissipative) dynamics of the mean field quantum Heisenberg
spin glass model in the large-$N$ limit. 
We have considered a particular type of initial condition which corresponds to the physical
situation in which, at $t=0^{-}$ the system is at equilibrium at
infinite temperature and at $t=0^{+}$ becomes coupled 
to  thermal bath in equilibrium at temperature $T$. This corresponds to an extremely fast quench 
from very high temperature. Depending on the value of $T$, the system
has a very different long-time behavior. 

At high temperature
the system relaxes, after a finite equilibration time, inside the paramagnetic state.
In this stationary regime the system is at equilibrium and the fluctuation dissipation
relation holds. When the system is quenched below a certain critical temperature $T_{d}$,  
which depends on the values of the spin and the system-bath coupling, it never reaches
an equilibrium regime. At large times two time-sectors can be identified for the behavior 
of the correlation $C_{S}(t,t')$ and response $R_{S}(t,t')$. When $t$ and $t'$ are very large, 
but their difference remains of the order of one, the systems reaches a pseudo-equilibrium
regime in which the QFDR is verified. However on a larger timescale, diverging with the 
age of the system ($t-t'\propto t,t'$), there is a secondary relaxation called aging.
In this regime the quantum fluctuation dissipation relation is violated and the correlation 
and the response are related by a generalization of the {\sl
classical} fluctuation dissipation relation 
characterized by an effective temperature different from the bath temperature.

Moreover we have also studied the role of the bath.
First, we have taken a linear coupling of the spin to the bath, we have developed to the second order in
the coupling constant and integrated out the bath spins. In this
way we have found a generalization 
of the Feynman-Vernon influence functional \cite{FeynmanVernon} for spins
in which the properties of the bath enters 
only through its susceptibility. All the numerical study has been done in this case. However, 
we have also considered a more general type of bath and we have shown
that a ``simple'' one turns out to be a Kondo bath. 
We have extended the analytical study to this case and shown that the 
previous dynamical scenario continues to hold. Furthermore we have unveiled
that the way we have followed previously 
to treat the system-bath coupling can be recover as a limiting case of a Kondo Bath.
Finally, we have also verified numerically that, as far as we can go
increasing the coupling to the bath 
in (\ref{KeldyshAction1Site}),
the dynamical transition remains of second order (by this we means
that the asymptotic dynamical energy  
is continuous) and the critical temperature does not vanish.

The most striking features of the low temperature out of equilibrium
dynamics are the aging phenomenon and  
the generalization of the fluctuation dissipation relation out of equilibrium. It has been shown for
spherical spins \cite{CugliandoloLozano} and for rotors \cite{ChamonRotors,ChamonRotorsLettre} that
 a generalization of the classical fluctuation dissipation relation
holds in the aging regime. In this paper we have shown  
that this is the case also for models with a non trivial spin
algebra. These results seem to suggest that, 
except for the renormalization of the coefficients of the dynamical
equations, the aging regime is not  
affected by quantum fluctuations and the aging systems behaves classically in their slow evolution.
 But is this always true? Is it not possible to find ``a quantum system which ages coherently'' ? 
Since in general, the decoherence time  
is finite and the aging regime takes place in the large time limit, a
classical aging regime is always 
 expected to set in at large enough time. However there is an important
case in which this naive argument  
may fail. Near a quantum critical point the decoherence time diverges,
therefore it could be possible  
that at very large times (larger than the time on which the system enters in the asymptotic regime
and than the characteristic timescale of the TTI-regime), but still
lower than the decoherence time, the 
system {\it ages coherently}. We could not address this very interesting 
question for the quantum Heisenberg spin glass analyzed
in this paper : the technical reason is that its  
quantum critical point is rather pathological since it corresponds to
a vanishing spin size. Hence, another type  
of model with a less singular quantum critical point has to be
studied. Work is in progress in this direction \cite{WorkInProgress}.  

\acknowledgments

Both authors are supported by the Center of Material Theory, Rutgers University.
We also thank NSF DMR 0096462 and the Rutgers Computational Grid
for support for the numerical computations.

\appendix

\section{From real time to imaginary time}\label{AppendiceMatsubara}
In this appendix, we give explicit formulas for doing the Wick
rotation to imaginary time in equilibrium. 
Let us define : 
\begin{equation}\label{DefCrochetG}
[G] (t) \equiv \left\{
         {  \begin{aligned} 
              & G_{++} (t) \text{ for  } t>0 \\
              & G_{--} (t) \text{ for  } t<0 
           \end{aligned}
         }
\right.
\end{equation}
We note that this function 
has a simple expression in terms of the spectral density (using
equilibrium FDT) :
\begin{align*}\label{gcro2}
[G] (t)&= 
 \begin{cases}
  & i \int d\epsilon \rho (\epsilon ) \check n_{B} (\epsilon ) e^{-i\epsilon t}
  \text{ for bosons } \\
 & -i \int d\epsilon \rho (\epsilon ) \check n_{F}(\epsilon ) e^{-i\epsilon t}
  \text{ for fermions }
 \end{cases} 
\\
[\check G] (t)&= 
 \begin{cases}
  & -i \int d\epsilon \rho (\epsilon )  n_{B}(\epsilon ) e^{i\epsilon t}
  \text{ for bosons } \\
 & i \int d\epsilon \rho (\epsilon )  n_{F}(\epsilon ) e^{i\epsilon t}
  \text{ for fermions }
 \end{cases} 
\end{align*}
where $n_{B}(\epsilon )$  and $n_{F}(\epsilon )$  are the Fermi and
Bose function respectively and we use the notation $\check f (x)\equiv f (-x)$.
Thus $[G]$ is analytic in $t$, and we have the relation :
\begin{align}
[G] (-i\tau ) &= i G (\tau ) \qquad   0<\tau <\beta  \\
[\check G] (-i\tau ) &=
\begin{cases}
&i G (\beta -\tau ) \qquad  0<\tau <\beta \qquad 
                    \text{for bosons }\\
&-i G (\beta -\tau ) \qquad  0<\tau <\beta \qquad 
                    \text{for fermions } 
\end{cases}
\end{align}
where the Matsubara Green function is defined by
\cite{NegeleOrlandBook} :
\begin{align}\label{App.DefGtau}
G (\tau ) &\equiv   - \moy{T b (\tau)b^{\dagger} (0)} = 
\int d\epsilon \, \rho (\epsilon )  \check n_{B} (\epsilon )
e^{-\epsilon \tau } 
 \qquad  \text{for
bosons } \\
G (\tau ) &\equiv   - \moy{T f (\tau)f^{\dagger} (0)} =
-\int d\epsilon \, \rho (\epsilon )  \check n_{F} (\epsilon )
e^{-\epsilon \tau } 
  \qquad  \text{for
fermions } 
\end{align}
The same formula also holds for the self-energy.

Using this result, we find the imaginary time equations in the paramagnetic
state : 
\begin{subequations}\label{App.EqMatsubara}
\begin{align}
(G^{-1}) (i\nu_{n}) &=i\nu_{n} + \lambda - \Sigma (i\nu_{n}) \\
\Sigma(\tau ) &=\ G(\tau) 
\biggl (J_{H}^{2} G (\tau )G (-\tau ) + J_{B}^{2} \chi^{0} (\tau )
\biggr)\\
G (\tau=0^{-} ) &= - S
\end{align}
\end{subequations}
They are a slight generalization of Eq. (5) of \cite{SGlong},
including the bath.

Similarly, in the glassy phase, using the same technique, we find
Eqs. (31) of \cite{SGlong} with $\Theta= \frac{1}{\sqrt{3}}$, which
corresponds to the marginality criterion, as explained in \cite{SGlong}.

\section{Numerical solution}\label{AppendiceNumerique}

In this appendix, we provide some details about the numerical
solution of our main equations (\ref{EqsGenerales}).
In order to compute $R$ and $K$ on the domain $0<t'<t$, we use the
causality of the equations (\ref{EqsGenerales}) :  
in order to compute the function in $(t,t')$  we only need
the knowledge of the functions at previous times, so we can construct the
functions step by step in time along the $t$ direction. This structure of the equations is 
general for classical or quantum spin glass dynamical problems.
(See e.g. \cite{CugliandoloLozano}).
We have to solve a set of coupled differential equations in $t$.
However, in this problem the situation is more complicated, since we
have first to compute an unphysical bosonic function, which oscillates
a lot. A naive algorithm is to compute the derivatives a
each point $(t,t')$ for a fixed $t$, and extrapolate using a first
order Taylor expansion. 
However for  numerical integration of ordinary differential
equations (ODE), this method is not recommended  (See e.g. \cite{NumericalRecipes}),
since one needs a very tiny mesh size to obtain accurate result.
In our case, we found that this simple algorithm does not give any
good result for a reasonable computational cost, contrary to simpler
models studied previously (e.g. classical $p-$spins models). 

Hence, we used a modified procedure, inspired by the Stoer-Burlish
algorithm for (ODE) : let us assume that we have computed
 the functions until time $t$ and we want them at time $t+\delta$
where $\delta$ is our mesh size. We cut this step into $N$ parts, and
compute the functions for $t + i \delta /N$ for all $t'$ and $1\leq
i\leq N$, using the modified midpoint method \cite{NumericalRecipes}.
We then obtain the functions at $t+\delta$, for various $N$ and all
$t'$ and we extrapolate the result to $N\rightarrow
\infty$. Typically, we use 3 or 4 values of $N$  among $\{4,8,16,32\}$.
The integrals are computed using either a trapezoidal or a Simpson
formula. It is important to notice that the structure of the equations
(\ref{EqsGenerales}) implies that we do not need to keep the
intermediate point after the $t+\delta$ have been computed.
As explained in the text, the dynamical equations conserve the
constraint, which is  automatically satisfied in the time evolution
: it is also important for the stability of the algorithm
that its discrete implementation of Eqs. (\ref{EqsGenerales})
respects this conservation exactly.

\input{SG.bbl}

\end{document}

%% file: SG.bbl
\newcommand{\PRB}{Phys. Rev. B}\newcommand{\PRL}{Phys. Rev. Lett}\newcommand{\NPB}{Nucl. Phys.}\newcommand{\RMP}{Rev. Mod. Phys.}\newcommand{\ADV}{Adv. Phys.}

%% file: SG.bbl
\begin{thebibliography}{79}
\bibitem{Note_Keldysh}\bibinfo{note}{By ``Keldysh action'' ${\cal S}$, we mean that $\moy{A} = \int {\cal D}S
\, A \exp ({\cal  S})$.}.
\bibitem{Note_FiniteTimescales}\bibinfo{note}{Finite timescales means timescales which do not
                                  diverge with the size of the system, i.e. one has to take
                                  the thermodynamic  
                                  limit before the long time limit.}.
\bibitem{Note_EA}\bibinfo{note}{The Edwards-Anderson parameter for the bosonic functions should be
                      purely imaginary since  $\lim_{t-{t'}\rightarrow \infty}\lim_{{t'}
                      \rightarrow \infty}
                      K(t,{t'})=-i\moy{b}\moy{b^{\dagger }}$}.
\end{thebibliography}
